\documentclass[aps,superscriptaddress,nofootinbib,eqsecnum,11pt]{revtex4}
\usepackage[english]{babel}
\usepackage[utf8]{inputenc}
\usepackage{graphicx}   
\usepackage{slashed}
\usepackage{verbatim}   
\usepackage{color}      
\usepackage{subfigure}  
\usepackage{multirow}
\usepackage{hyperref}   
\usepackage{float}
\usepackage{epsfig,rotating}
\usepackage{amsmath,amssymb}
\usepackage{dsfont}
\restylefloat{table}
\numberwithin{equation}{section}

\newcommand{\vx}{\vec{x}}
\newcommand{\vp}{\vec{p}}
\newcommand{\vq}{\vec{q}}
\newcommand{\vk}{\vec{k}}

\newcommand{\be}{\begin{equation}}
\newcommand{\ee}{\end{equation}}
\newcommand{\bea}{\begin{eqnarray}}
\newcommand{\eea}{\end{eqnarray}}

\begin{document}
\title{ Imprint of entanglement entropy in the power spectrum of inflationary fluctuations.}

\author{Daniel Boyanovsky}
\email{boyan@pitt.edu} \affiliation{Department of Physics and
Astronomy, University of Pittsburgh, Pittsburgh, PA 15260}
\date{\today}

\begin{abstract}
If the inflaton couples to other degrees of freedom that populate the post-inflationary stage, such coupling modifies the dynamics of the inflaton \emph{during} inflation. We consider light fermions Yukawa coupled to the inflaton as ``unobserved'' degrees of freedom  integrated out of the total density matrix. Tracing out these degrees of freedom yields a \emph{mixed} density matrix whose time evolution is described by an effective field theory. We show that the coupling leads to profuse fermion pair production for super-Hubble inflaton fluctuations which lead to the \emph{growth of entanglement entropy during inflation}. The power spectrum of inflaton fluctuations features scale invariance violations $\mathcal{P}(k) = \mathcal{P}_0(k)\,\,\exp\{8\,\xi_k\}$
with corrections to the \emph{index and its  running directly  correlated with the entanglement entropy}: $S_{vN} = - \sum_{k} \Big[ \ln(1-\xi_k) + \frac{\xi_k\,\ln(\xi_k)}{1-\xi_k} \Big]$. For super-Hubble fluctuations we find
$\xi_k = -\frac{Y^2}{48\pi^2}\Big\{2\,N_T\,\ln(k/k_f) + \ln^2(k/k_f)\Big\}$ with $Y$ the Yukawa coupling, $N_T$ the total number of e-folds during inflation, and $k_f$ a ``pivot'' scale corresponding to the mode that crosses the Hubble radius at the end of inflation.

\end{abstract}


\maketitle

\section{Introduction, Motivation and goals}\label{sec:intro}

  The main predictions of inflationary cosmology are supported by observations of the cosmic microwave background (CMB) anisotropies with unprecedented accuracy by the WMAP\cite{wmap,spergel} and PLANCK\cite{planck} missions.   A simple paradigm of inflationary cosmology describes the inflationary stage as dominated by the dynamics of a   scalar field, the inflaton, slowly rolling down a potential landscape leading to a nearly de Sitter inflationary epoch\cite{guth,linde}. During this period (adiabatic) cosmological perturbations   are generated by  quantum fluctuations that freeze when their wavelengths become larger than the Hubble radius  with a nearly scale invariant power spectrum\cite{mukhanov,bran}. Upon re-entering the Hubble radius during the matter dominated era, these fluctuations  provide the seeds for  structure formation.

  Typical  models of inflation invoke one scalar field, the inflaton, yielding adiabatic perturbations, including other scalar fields   generically yield a small component of isocurvature (entropy) perturbations that are severely constrained by CMB observations\cite{wmap,spergel,planck}.    The interactions between the inflaton field and other fields describing degrees of freedom that   populate a post-inflation, radiation dominated era, such as those present in the Standard Model,  are usually considered within the realm of reheating post-inflation\cite{reheat1,reheatrev}. However, if the inflaton interacts with other degrees of freedom, these interactions do not suddenly ``switch-on'' after the inflationary stage, but must be present even \emph{during} inflation. Hence,  a logical conclusion  is that,  if the theory of reheating is relevant to describe the post-inflationary cosmology,   the degrees of freedom excited during this stage will also be coupled to the inflaton \emph{during} inflation. From this perspective, scalar field-driven inflation should be understood as an \emph{effective field theory} emerging after tracing out, or coarse graining, these ``unobserved'' degrees of freedom that are not \emph{directly} involved in the generation of the cosmological perturbations that seed the temperature anisotropies.

  Interacting quantum fields in a de Sitter (or nearly de Sitter) space time have been the focus of several important studies\cite{woodardcosmo}-\cite{polyakov} that pointed out the emergence of secular and infrared divergences associated with nearly massless fields in   inflationary cosmology. Previous studies have shown that loop contributions from ``spectator'' fields feature these secular or infrared divergences and may yield a time dependence of curvature perturbations in the super-Hubble limit\cite{kahya1,kahya2,covi}. An important framework to study effective field theories out of equilibrium is that of open quantum systems wherein the time evolution of a reduced density matrix, obtained by tracing over unobserved degrees of freedom, is determined by a quantum master equation\cite{breuer,zoeller,boyeff}. This approach has recently began to be implemented in cosmology\cite{burhol,boydensmat,hollowood,oshita,shandera,vennin,kanno} and shown to be equivalent to the non-equilibrium effective action that includes the influence action of the degrees of freedom that are traced over\cite{boyeff,boyeffcosmo1}. The influence of these ``unobserved'' degrees of freedom, including fermions\cite{boyfer} has been shown to lead to corrections to the power spectrum of inflaton fluctuations and violations of scale invariance\cite{boydensmat,boyeffcosmo1}. Recently    the interaction between the inflaton   and ``environmental'' fields has been studied within the framework of the quantum master equation to assess the discord, namely the effect of decoherence on inflationary correlations\cite{hollowood,oshita}. Taken together, these studies suggest  a relationship between corrections to the power spectrum from ``environmental'' fields and discord and decoherence as a consequence of interactions with ``unobserved'' (traced over) degrees of freedom.

  \vspace{1mm}

  \textbf{Motivation, goals and main results:}

  Our study is motivated by the following aspects: \textbf{i)}  The power spectrum of scalar perturbations is characterized by the index $n_s$ with the tilt\cite{wmap,planck} $(1-n_s)$ indicating (slight) violations of scale invariance, with the \emph{running} $\alpha_s= d\,n_s/d\,\ln(k)$ and \emph{running of the running} $\beta_s = d\,\alpha_s/d\,\ln(k)$ being higher order indicators of violations of scale invariance. In single field slow roll scenarios $1-n_s,\alpha_s,\beta_s$ follow a hierarchy in slow roll parameters. The analysis of the Planck collaboration\cite{planck}, however,  yields a value of $\beta_s$ surprisingly large, positive, and of the same order of but slightly larger than $\alpha_s$ that seems to be in tension with slow roll scenarios\cite{escudero,cabass}, although at the $\simeq 2 \sigma$ level. However, future surveys may tighten this bound\cite{kamion}.
  Values of $\alpha_s,\beta_s$ larger (and of different sign) than those predicted in the simple single field slow roll inflation can be obtained by allowing entropy perturbations\cite{bruck} or from contributions of other sources\cite{kimmo}.

  \textbf{ii)} If the inflaton is coupled to the degrees of freedom that describe the post-inflation radiation dominated phase, this coupling is also present \emph{during} inflation.   A corollary of the  results of  refs.\cite{boydensmat,boyeffcosmo1,hollowood,boyfer}, is that the interaction between the inflaton  and   ``unobserved'' degrees of freedom that are integrated out into an effective dynamics, yield   corrections to the power-spectrum of inflationary quantum fluctuations with violations of (near) scale invariance.   Remarkably, these corrections obtained in refs.\cite{boydensmat,boyeffcosmo1,boyfer,hollowood} can also be interpreted as a running   $\alpha_s$   determined by the interaction strength.

  \textbf{iii)}
  A recent study showed that integrating out (``unobserved'') degrees of freedom to yield an effective field theory implies a \emph{loss of information}, which is manifest as an \emph{entanglement entropy } of the effective field theory that determines the time evolution of the reduced density matrix\cite{boyinfo}.

   Our aim   is to assess whether, and how, the information loss and entanglement entropy encoded in the effective field theory resulting from tracing out the ``unobserved'' degrees of freedom\cite{boyinfo} is manifest or \emph{imprinted} in the corrections to the power spectrum of inflaton fluctuations. In other words,   we study the \emph{relationship} between the violations of scale invariance in the power spectrum induced by the coupling of the inflaton to the  unobserved  fields and the \emph{information loss} and \emph{entanglement entropy} arising from tracing over these degrees of freedom. If such a relationship can be unambiguously established, a measurement of $n_s;\alpha_s;\beta_s$ that departs from the predictions of single field slow roll \emph{may} be evidence of an underlying effective field theory description of inflation in which ``unobserved'' degrees of freedom yield corrections to observables.

  \vspace{1mm}

  \textbf{Main results:}

  Assuming that the scale of inflation $H$ is much larger than the weak scale, we consider the inflaton Yukawa coupled to fermions with masses $m_f \ll H$ as these are the most ubiquitous degrees of freedom of the standard model to which a real scalar field can couple directly.  We consider an initial factorized density matrix describing Bunch-Davies vacua for the inflaton and fermions, evolve this state in time in the interacting theory and trace the fermions out of the time evolved density matrix obtaining a reduced density matrix for the inflaton. We begin the study with a perturbative evaluation of the reduced density matrix. This approach makes evident that   the production of fermion-anti-fermion pairs \emph{kinematically entangled with inflaton fluctuations} leads to a mixed state upon tracing over the fermion pairs. The coefficients of the reduced density matrix reveal secular growing terms for super-Hubble  inflaton fluctuations, these are a consequence of profuse fermion pair production enhanced when the physical wavelength of inflaton fluctuations become super-Hubble. We obtain an preliminary \emph{estimate} of the entanglement entropy and its relation to the power spectrum  in the super-Hubble limit. We then obtain the one loop effective action upon integrating out the fermionic degrees of freedom and show that it yields the time evolution of the reduced density matrix from which we obtain the entanglement entropy and the power spectrum of inflaton fluctuations confirming the perturbative treatment. For the entanglement entropy we find
  \be S_{vN} = - \sum_{k} \Bigg\{ \ln(1-\xi_k) + \frac{\xi_k\,\ln(\xi_k)}{1-\xi_k} \Bigg\} \,, \label{entent}\, \ee where for super-Hubble modes and Yukawa coupling $Y$ we find
 \be \xi_k = -\frac{Y^2}{48\pi^2}\Big\{2\,N_T\,\ln(k/k_f) + \ln^2(k/k_f)\Big\}\,,\ee with $N_T$ the total number of e-folds during inflation, and $k_f$ a ``pivot'' scale corresponding to the mode that crosses the Hubble radius at the end of inflation.

  A dynamical renormalization group improvement yields for the inflaton power spectrum in the super-Hubble limit
  \be \mathcal{P}(k) = \frac{H^2}{4\pi^2}\,\, e^{8\,\xi_k}\,,  \label{pofketa}  \ee explicitly showing that the corrections to the power spectrum with scale invariance violations are \emph{directly} related to the entanglement entropy and information loss of the effective field theory.  The corrections to the scalar index and its running are given by
\be \delta n_s = - \frac{N_T\,Y^2}{3\pi^2} ~~;~~ \alpha_s = -\frac{Y^2}{6\pi^2}\,,  \ee with vanishing running of the running to leading order in $Y^2$.

\section{The model:}\label{sec:model}

We consider the model of an inflaton scalar field minimally coupled to gravity and Yukawa coupled to one Dirac fermionic degree of freedom  in a spatially flat de Sitter space time. Including Majorana fermions and/or more species is straightforward\cite{boyfer}.

In comoving
coordinates, the action is given by
\bea
S & = &\int d^3x \; dt \;  \sqrt{-g} \,\Bigg\{
\frac{1}{2}{\dot{\phi}^2}-\frac{(\nabla
\phi)^2}{2a^2}-\frac{1}{2} \, M^2  \phi^2   +
\overline{\Psi}  \Big[i\,\gamma^\mu \;  \mathcal{D}_\mu -m_f-Y \phi \Big]\Psi     \Bigg\}\,. \label{lagrads}
\eea

The Dirac $\gamma^\mu$ are the curved space-time $\gamma$ matrices
and the fermionic covariant derivative is given
by\cite{weinbergbook,BD,duncan,casta}
\bea
\mathcal{D}_\mu & = &  \partial_\mu + \frac{1}{8} \;
[\gamma^c,\gamma^d] \;  e^\nu_c  \; \left(D_\mu e_{d \nu} \right)
\cr \cr 
D_\mu e_{d \nu} & = & \partial_\mu e_{d \nu} -\Gamma^\lambda_{\mu
\nu} \;  e_{d \lambda} \nonumber
\eea
\noindent where the vierbein field $e^\mu_a$ is defined as
$$
g^{\mu\,\nu} =e^\mu_a \;  e^\nu_b \;  \eta^{a b} \; ,
$$
\noindent $\eta_{a b}$ is the Minkowski space-time metric. The curved space-time  matrices $\gamma^\mu$ are given in terms of
the Minkowski space-time ones $\gamma^a$  by (greek indices refer to
curved space time coordinates and latin indices to the local
Minkowski space time coordinates)
$$
\gamma^\mu = \gamma^a e^\mu_a \quad , \quad
\{\gamma^\mu,\gamma^\nu\}=2 \; g^{\mu \nu}  \; .
$$
For a  Friedmann Robertson Walker metric  in conformal time  the metric becomes
\be
g_{\mu\nu}= C^2(\eta) \;  \eta_{\mu\nu}
\quad , \quad C(\eta)\equiv a(t(\eta))\label{gmunu}
\ee

\noindent and $\eta_{\mu\nu}=\textrm{diag}(1,-1,-1,-1)$ is the flat
Minkowski space-time metric and for exact de Sitter space-time

 \be C(\eta) = -\frac{1}{H\eta}\,. \label{CdS}\ee

  In conformal time the vierbeins $e^\mu_a$ are given by
\be
 e^\mu_a = C^{-1}(\eta)\; \delta^\mu_a ~~;~~ e^a_\mu = C(\eta) \; \delta^a_\mu
\ee
\noindent and the Dirac Lagrangian density simplifies to
\be \label{ecdi}
\sqrt{-g} \; \overline{\Psi}\Big(i \; \gamma^\mu \;  \mathcal{D}_\mu
\Psi -m_f-Y\phi \Big)\Psi  =
(C^{\frac{3}{2}}\overline{\Psi}) \;  \Big[i \;
{\not\!{\partial}}-(m_f+Y\phi) \; C(\eta) \Big]
\left(C^{\frac{3}{2}}{\Psi}\right)
\ee
\noindent where $i {\not\!{\partial}}=\gamma^a \partial_a$ is the usual Dirac
differential operator in Minkowski space-time in terms of flat
space time $\gamma^a$ matrices.

Introducing the conformally rescaled fields
\be C(\eta) \phi(\vx,t) = \chi(\vx,\eta)~~;~~ C^{\frac{3}{2}}(\eta){\Psi(\vx,t)}= \psi(\vx,\eta)\,, \label{rescaledfields}\ee
and neglecting surface terms, the action becomes
   \be  S    =
  \int d^3x \; d\eta \, \Big\{\mathcal{L}_0[\chi]+\mathcal{L}_0[\psi]+\mathcal{L}_I[\chi,\psi] \Big\} \;, \label{rescalagds}\ee
  where
  \bea \mathcal{L}_0[\chi] & = & \frac12\left[
{\chi'}^2-(\nabla \chi)^2- \mathcal{M}^2 (\eta) \; \chi^2 \right] \,, \label{l0chi}\\
\mathcal{L}_0[\psi] & = & \overline{\psi} \;  \Big[i \;
{\not\!{\partial}}+ \frac{m_f}{H\eta}    \Big]
 {\psi}  \,,\label{l0psi}\\ \mathcal{L}_I[\chi,\psi] & = & -Y\chi :\overline{\psi}\,\psi: \; , \label{lI}\eea where we have normal ordered the interaction in the interaction picture of free fields, and
 \be
\mathcal{M}^2 (\eta)  = \Big[\frac{M^2}{H^2}-2\Big]\frac{1}{\eta^2} \,. \label{masschi2}\ee

In the non-interacting case $Y =0$ the Heisenberg equations of motion for the spatial Fourier modes of wavevector $\vec{k}$ for the conformally rescaled scalar field are
\be
  \chi''_{\vk}(\eta)+
\Big[k^2-\frac{1}{\eta^2}\Big(\nu^2_\chi -\frac{1}{4} \Big)
\Big]\chi_{\vk}(\eta)  =   0\,      \label{chimodes}\ee
where
\be   \nu^2_{\chi} = \frac{9}{4}-  \frac{M^2}{H^2}  \,.
\label{nuchi} \ee We consider a light inflaton   field     with
$   M^2/H^2  \ll 1  $   consistently with a nearly scale invariant power spectrum.

The Heisenberg fields are quantized in a comoving volume $V$ as
\be
\chi(\vx,\eta)   =   \frac{1}{\sqrt{V}}\,\sum_{\vq} \Big[a_{\vq}\,g(q,\eta)\,e^{i\vq\cdot\vx}+ a^\dagger_{\vq}\,g^*(q,\eta)  \,e^{-i\vq\cdot\vx}\Big] \,.\label{chiex}  \ee

We choose Bunch-Davies conditions for the scalar fields, namely
\be a_{\vq} |0\rangle_{\chi} =0 \,. \label{bdvac}\ee
and
\be
  g(q,\eta)= \frac{1}{2}\,e^{i\frac{\pi}{2}(\nu_\chi+\frac{1}{2})}\,\sqrt{-\pi\,\eta}~H^{(1)}_{\nu_\chi}(-q\eta)
 \label{gqeta}\,. \ee For $M^2/H^2 \ll 1$ corresponding to $\nu_{\chi} \simeq 3/2$ the mode functions simplify to
 \be g(q,\eta) = \frac{e^{-ik\eta}}{\sqrt{2k}}\Big[1- \frac{i}{k\eta} \Big]\,. \label{caso32}\ee

  The Dirac equation for Fermi fields becomes
\be  \Big[i \;
{\not\!{\partial}}- M_\psi(\eta)    \Big]
 {\psi}  = 0 ~~;~~M_\psi(\eta) = -   \frac{m_f}{H\eta} \,.\label{diraceqn}\ee
For Dirac fermions the solution $ \psi({\vec x},\eta) $ is expanded  as
\be
\psi (\vec{x},\eta) =    \frac{1}{\sqrt{V}}
\sum_{\vec{k},\lambda}\,   \left[b_{\vec{k},\lambda}\, U_{\lambda}(\vec{k},\eta)\,e^{i \vec{k}\cdot
\vec{x}}+
d^{\dagger}_{\vec{k},\lambda}\, V_{\lambda}(\vec{k},\eta)\,e^{-i \vec{k}\cdot
\vec{x}}\right] \; ,
\label{psiex}
\ee
where the spinor mode functions $U,V$ obey the  Dirac equations
\bea
\Bigg[i \; \gamma^0 \;  \partial_\eta - \vec{\gamma}\cdot \vec{k}
-M_\psi(\eta) \Bigg]U_\lambda(\vec{k},\eta) & = & 0 \,,\label{Uspinor} \\
\Bigg[i \; \gamma^0 \;  \partial_\eta + \vec{\gamma}\cdot \vec{k} -M_\psi(\eta)
\Bigg]V_\lambda(\vec{k},\eta) & = & 0 \,.\label{Vspinor}
\eea

We choose to work with the standard Dirac representation of the (Minkowski) $\gamma^a$ matrices.

It proves
convenient to write
\bea
U_\lambda(\vec{k},\eta) & = & \Bigg[i \; \gamma^0 \;  \partial_\eta -
\vec{\gamma}\cdot \vec{k} +M_\psi(\eta)
\Bigg]f_k(\eta)\, \mathcal{U}_\lambda \,,\label{Us}\\
V_\lambda(\vec{k},\eta) & = & \Bigg[i \; \gamma^0 \;  \partial_\eta +
\vec{\gamma}\cdot \vec{k} +M_\psi( \eta)
\Bigg]h_k(\eta)\,\mathcal{V}_\lambda \,,\label{Vs}
\eea
\noindent with $\mathcal{U}_\lambda;\mathcal{V}_\lambda$ being
constant spinors\cite{boydVS,baacke} obeying
\be
\gamma^0 \; \mathcal{U}_\lambda  =  \mathcal{U}_\lambda
\label{Up} \qquad , \qquad
\gamma^0 \;  \mathcal{V}_\lambda  =  -\mathcal{V}_\lambda
\ee
The mode functions $f_k(\eta);h_k(\eta)$ obey the following
equations of motion
\bea \left[\frac{d^2}{d\eta^2} +
k^2+M^2_\psi(\eta)-i \; M'_\psi(\eta)\right]f_k(\eta) & = & 0 \,, \label{feq}\\
\left[\frac{d^2}{d\eta^2} + k^2+M^2_\psi(\eta)+i \; M'_\psi(\eta)\right]h_k(\eta)
& = & 0 \,.\label{geq}
\eea
We choose Bunch-Davies boundary conditions for the solutions, namely
\be f_k(\eta) ~~ {}_{\overrightarrow{-k\eta \rightarrow \infty}}~~ e^{-ik\eta}~~;~~ h_k(\eta) ~~{}_{\overrightarrow{-k\eta \rightarrow \infty}}~~ e^{ik\eta} \,, \label{bdfketa}\ee which leads to the choice
\be h_k(\eta)= f^*_k(\eta)\,, \label{choice} \ee  and $f_k(\eta)$ is a solution of
\be \left[\frac{d^2}{d\eta^2} +
k^2+
\frac{1}{\eta^2}\Big[\frac{m^2_f}{H^2}-i\frac{m_f}{H}\Big]\right]f_k(\eta)   =   0 \,. \label{eqnfketa}\ee   We find
\be f_k(\eta) = \sqrt{\frac{-\pi k \eta}{2}}\,\,e^{i\frac{\pi}{2}(\nu_\psi+1/2)}\,\,H^{(1)}_{\nu_\psi}(-k\eta)~~;~~ \nu_\psi = \frac{1}{2}+i\frac{m_f}{H}\,. \label{fketasolu}\ee The sub-Hubble limit $(-k\eta) \rightarrow \infty$ of these modes is given by (\ref{bdfketa}) whereas these modes feature a purely oscillatory  super-Hubble behavior\cite{boyfer}.   The important aspect, however, is that the amplitude of the mode functions remains bound and of order unity for super-Hubble wavelengths.

Under the assumption that the scale of inflation is much larger than the weak scale and that the fermionic degrees of freedom represent those of the Standard Model, it follows that $H \gg m_f$, leading to
\be f_k \simeq e^{-ik\eta} \,,\label{lightferlim}\ee in contrast,    nearly massless  $M\ll H$ minimally coupled scalar fields feature a growing mode in the super-Hubble limit ($k\eta \ll 1$) with
\be g(k,\eta) \propto \frac{ 1}{k^{3/2}\eta}\,, \label{grow}\ee  which results in amplification and classicalization of super-Hubble fluctuations\cite{polarski}.

\section{Entanglement entropy and power spectrum: a  perturbative argument.}\label{sec:pert}
Before we study the time evolution of the reduced density matrix via the effective action, we analyze the emergence of an entangled state between inflaton and fermionic degrees of freedom in perturbation theory. The aim of  this section is  to  provide a simple physical understanding of the emergence of the entanglement entropy and its relation to the power spectrum, along with a \emph{preliminary} estimate of its value.  The results of this section  must be taken as \emph{indicative}, and as a guide to the physical processes involved.  The next sections provide a more technically   detailed  and firmer derivation of the reduced density matrix and entanglement entropy from the effective action.

 Consider an initial state corresponding to the Bunch-Davies vacuum for both the  inflaton and fermions, namely
\be |\Psi(\eta_0)\rangle = |0\rangle_\chi\otimes |0\rangle_\psi \,.\label{inistate} \ee In the Schroedinger picture the time evolution of this state is given by
\be |\Psi(\eta )\rangle = U(\eta;\eta_0) \,|\Psi(\eta_0)\rangle \label{psifoeta}\ee where $U(\eta;\eta_0)$ is the unitary time evolution operator obeying
\be i \frac{d}{d\eta}\,U(\eta;\eta_0) = H(\eta)\,  U(\eta;\eta_0)~~;~~ U(\eta_0;\eta_0) =1 \,, \label{Uofeta}\ee where $H(\eta)= H_0(\eta)+H_i(\eta)$ is the total Hamiltonian, and $H_0(\eta)$,$H_i(\eta)$ are the free field and interaction Hamiltonian respectively, with $H_0(\eta)$ depending  explicitly on $\eta$ through the mass terms. The \emph{reduced} density matrix for the inflaton field $\chi$ is obtained by performing the trace over the fermionic degrees of freedom of the full density matrix. It  is given by
\be \rho^r_{\chi}(\eta) = \mathrm{Tr}_{\psi}\big(|\Psi(\eta)\rangle\langle\Psi(\eta)|\big)\,. \label{redchi}\ee

 Entanglement between the inflaton and fermionic degrees of freedom resulting from their interaction and time evolution is best studied in the interaction picture. The unitary time evolution operator in absence of interaction (free fields) $U_0(\eta;\eta_0)$ obeys
 \be i \frac{d}{d\eta}\,U_0(\eta;\eta_0) = H_0(\eta)\,  U_0(\eta;\eta_0)~~;~~ U_0(\eta_0;\eta_0) =1 \,, \label{Uzeroeta}\ee The quantum state in the interaction picture evolves in time as
 \be |\Psi(\eta)\rangle_I = U_I(\eta;\eta_0) |\Psi(\eta_0)\rangle \label{psinte} \ee where the unitary time evolution operator in the interaction picture $U_I(\eta;\eta_0) = {U_0}^{-1}(\eta;\eta_0) \,U(\eta;\eta_0) $ obeys
\be i \frac{d}{d\eta}\,U_I(\eta;\eta_0) = H_I(\eta)\,  U_I(\eta;\eta_0)~~;~~ U_I(\eta_0;\eta_0) =1 \,, \label{Uofetaip}\ee where
\be H_I(\eta) = Y\,\int d^3 x \,\chi(\vec{x},\eta) \, :\overline{\psi}(\vec{x},\eta)\,\psi(\vec{x},\eta)\,: \label{HIint}\ee is the interaction Hamiltonian in the interaction picture,  and $\chi,\psi$ are given by the free field expansions (\ref{chiex},\ref{psiex}) respectively.

Up to second order  in $Y$ we obtain
\be |\Psi(\eta)\rangle_I   = |\Psi(\eta_0)\rangle + |\Psi^{(1)}(\eta)\rangle +|\Psi^{(2)}(\eta)\rangle  + \cdots \label{timeevol} \ee where $|\Psi(\eta_0)\rangle$ is given by (\ref{inistate}) and
\bea |\Psi^{(1)}(\eta)\rangle  & =  &  -i\int^{\eta}_{\eta_0} H_I(\eta_1) d\eta_1 \,|\Psi(\eta_0)\rangle \label{state1}\\
|\Psi^{(2)}(\eta)\rangle  & =  &  (-i)^2\int^{\eta}_{\eta_0} \int^{\eta_1}_{\eta_0}\,H_I(\eta_1)H_I(\eta_2) \, d\eta_1 \, d\eta_2 ~~|\Psi(\eta_0)\rangle \nonumber \\ &  = &   -i\int^{\eta}_{\eta_0} H_I(\eta_1)\,|\Psi^{(1)}(\eta_1)\rangle ~  d\eta_1 \,.\label{state2}\eea

We find
 \be |\Psi^{(1)}(\eta)\rangle = \sum_{\vec{k},\vec{q}} M^{(1)}_{\lambda,\lambda'}(\vec{k},\vec{q};\eta)\,|1_{\vec{k}}\rangle_{\chi}\otimes |1_{\vec{q},\lambda};\overline{1}_{\vec{p},\lambda'}\rangle_{\psi}   ~~;~~ \vec{p} = -\vec{q}-\vec{k}\,, \label{Psi1t}\ee  where the matrix element is given by
 \be M^{(1)}_{\lambda,\lambda'}(\vec{k},\vec{q};\eta) = \frac{Y}{\sqrt{V}}\,\int^{\eta}_{\eta_0} d\eta_1  g^*(k,\eta_1)\, \overline{U}_\lambda(\vec{q},\eta_1) \, V_{\lambda'}(\vec{p},\eta_1)\,. \label{mtxele}\ee

\begin{figure}[ht!]
\begin{center}
\includegraphics[height=3in,width=3in,keepaspectratio=true]{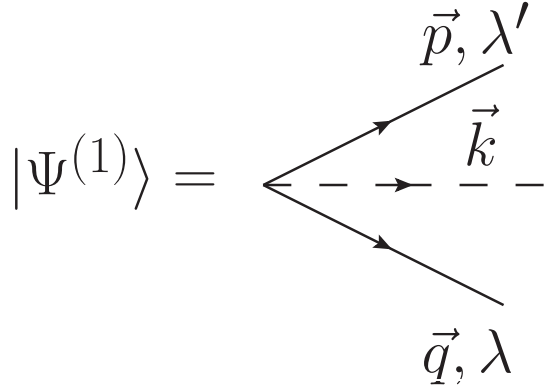}
\caption{ The first order state $\Psi^{(1)}$ is a multiparticle state with an inflaton (dashed line) kinematically entangled with a fermion-antifermion pair (solid lines) with $\vec{k} = -\vec{p}-\vec{q}$.  }
\label{fig:firstorder}
\end{center}
\end{figure}

 This state depicted in fig. (\ref{fig:firstorder}), it is recognized as  an   \emph{entangled} multiparticle state of the inflaton field and fermion-antifermion \emph{pairs}.

  An important aspect of the matrix element $M^{(1)}_{\lambda,\lambda'}(\vec{k},\vec{q};\eta)$  is that it \emph{grows} with conformal time for super-Hubble inflaton modes: as a consequence of the growing mode (\ref{grow})  for wavevectors that become super-Hubble at a time $\eta^* \approx -1/k$  the time integral in (\ref{mtxele}) yields a contribution $\propto Y \ln(\eta/\eta^*)$.

\begin{figure}[ht!]
\begin{center}
\includegraphics[height=3in,width=3.5in,keepaspectratio=true]{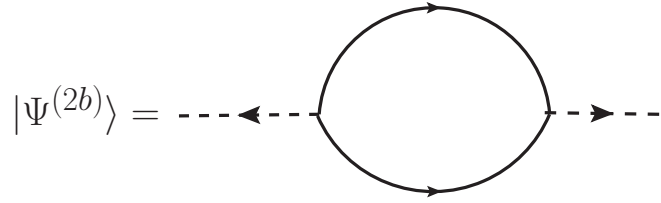}
\caption{ The second order state $\Psi^{(2b)}$ the solid line is a fermion-antifermion self-energy loop. The dashed lines are inflaton states.  }
\label{fig:secondorder}
\end{center}
\end{figure}

 In second order, there are several contributions obtained from the second equality in eqn. (\ref{state2}), however, only \emph{two} contribute to the reduced density matrix: i) annihilate all particles from $|\Psi^{(1)}(\eta)$ returning to the vacuum state  $|\Psi(\eta_0)\rangle$, ii) create another $\chi$-particle annihilating the fermion-antifermion pair in the state $|\Psi^{(1)}(\eta)\rangle$, yielding
 \be |\Psi^{(2)}(\eta)\rangle = M^{(2)}_a (\eta)\,|\Psi(\eta_0)\rangle+ \sum_{\vk} M^{(2)}_b (k;\eta)\,|1_{\vk};1_{-\vk}\rangle_\chi\, \otimes |0\rangle_\psi \,.\label{psi2}\ee
 The contribution to the second order state $|\Psi^{(2b)}\rangle$ with matrix element $M^{(2)}_b (k;\eta)$ is shown in fig. (\ref{fig:secondorder}), this is a one-loop self energy diagram.

 The matrix elements involve another wavefunction $g^*(\eta)$ and another time integral implying an extra logarithmic growth for super-Hubble wavevectors leading to a behavior
 \be  M^{(2)}_a (\eta) \propto Y^2 \ln^2 (\eta/\eta^*)~~;~~ M^{(2)}_b (k;\eta) \propto Y^2 \ln^2(\eta/\eta^*) \,. \label{mtxels2nd}\ee Therefore, the reduced density matrix in the interaction picture up to second order is given by
 \bea \rho^r_\chi(\eta) = \mathrm{Tr}_{\psi} \Big(|\Psi(\eta) \rangle \langle \Psi(\eta)| \Big)  &\simeq& \rho_0(\eta)\,\big(|0\rangle_\chi {}_\chi\langle 0|\big) + \sum_{\vk} \Big[\rho_1(\vk,\eta)\,\big(|1_{\vk}\rangle_\chi {}_\chi\langle 1_{\vk}|\big) \nonumber \\ & + & \rho_2(\vk,\eta)\,\big(|0\rangle_\chi {}_\chi\langle 1_{\vk};1_{-\vk}|\big)+  \rho^*_2(\vk,\eta)\,\big(| 1_{-\vk};1_{\vk}\rangle_\chi {}_\chi\langle0|\big)\Big] \,, \label{rho2} \eea where
 \be  \rho_0(\eta)   \simeq   \big[1+ 2 \, \mathrm{Re} M^{(2)}_a (\eta)  \big] ~~;~~
 \rho_1(\vk,\eta)   \propto   |M^{(1)}|^2 \propto Y^2 \ln^2(\eta/\eta^*)~~;~~ \rho_2(\vk,\eta) \propto M^{(2)}_b(\eta) \propto Y^2 \ln^2(\eta/\eta^*)\,. \label{rhos}\ee The secular growth of the matrix elements arising from super-Hubble wavelengths of inflaton fluctuations imply the profuse production of single quanta of  the inflaton   \emph{kinematically entangled}   and correlated with fermion pairs. That the reduced density matrix (\ref{rho2}) describes a mixed state can be understood from an argument closely related to that in ref.\cite{boyinfo}. \emph{If} (\ref{rho2}) were a pure state, it could be written as
 \be \rho^r_\chi(\eta) = |\alpha(\eta) \rangle \langle \alpha(\eta)| \,, \label{purerho}\ee where the state $|\alpha(\eta)\rangle$, up to second order in the coupling must be generically a superposition of the vacuum, single particle and correlated pair states as it must be obtained in a second order expansion with the interaction (\ref{lI}).  Therefore such state should be of the form
 \be |\alpha(\eta)\rangle  = \alpha_0(\eta) \,|0\rangle_\chi + \sum_{\vec{k}}\,\Big[ \alpha_1(\vk,\eta)\,|1_{\vk}\rangle_{\chi} + \alpha_2(\vk,\eta) |1_{\vk};1_{-\vk}\rangle_\chi+ \cdots \Big] \,,  \label{alfastat}\ee where $\alpha_0 \simeq \mathcal{O}(Y^0) + \mathcal{O}(Y^2)~;~ \alpha_1 \simeq \mathcal{O}(Y)~;~ \alpha_2 \simeq \mathcal{O}(Y^2)$. Comparing (\ref{purerho}) to (\ref{rho2}) we find that (\ref{purerho}) features terms first order in $Y$ of the form $\simeq |0\rangle_\chi~ _{\chi}\langle 1_{\vk}|$ and hermitian conjugate, as well as terms of the form $|1_{\vk}\rangle_\chi ~ _{\chi}\langle 1_{\vk'}|$ for $\vk \neq \vk'$. Neither of such terms are present in (\ref{rho2}). The main reason why these terms are not present in (\ref{rho2}) is because the single particle $\chi$ states are \emph{entangled} with fermion-antifermion pairs, in performing the trace over these degrees of freedom, each member of the pair in a ``bra'' must pair up with a similar state of same momentum and polarization $\lambda$ from a ``ket'' in the trace. Thus  we conclude that  the density matrix (\ref{rho2}) describes a \emph{mixed} state. The entanglement between the inflaton and fermionic states is responsible for the \emph{entanglement entropy} associated with this mixed state, which  is given by
 \be S = -\sum_{n} \lambda_n \ln( \lambda_n)\,, \label{Srho} \ee where $\lambda_n$ are the eigenvalues of the reduced density matrix.

 To leading order in $Y$, the eigenvalues in the single and two particle sectors are proportional to $\rho_0,\rho_1$ yielding an \emph{entanglement entropy}
 \be S \simeq -\sum_{\vk} \zeta_k(\eta) \ln\zeta_k(\eta)~~;~~\zeta_k(\eta) \propto Y^2\,\ln^2(-k\eta) \,. \label{entanS}\ee The growth of this entanglement entropy is a consequence of fermion pair production, which is enhanced when the physical wavelength of the inflaton fluctuation becomes super-Hubble. Tracing over the fermionic degrees of freedom leads to
 \emph{information loss} which is manifest as the entanglement entropy. As time evolves more inflaton modes become super-Hubble resulting in fermion-pair production, as more pairs are integrated out, more information is lost and entropy grows.

 The power spectrum of the original $\phi(\vec{x},\eta) = \chi(\vec{x},\eta)/C(\eta)$-field is given by
 \be \mathcal{P}_k(\eta) = \Big(\frac{k^3}{2\pi^2\, C^2(\eta)}\Big)\,\mathrm{Tr}_{\chi}\, \rho^r_{\chi}(\eta) \, \chi_{\vk}(\eta)\chi_{-\vk}(\eta) \label{Pofk} \ee where (see eqn. (\ref{chiex})) in the  interaction picture
 \be \chi_{\vk}(\eta) = a_{\vk} \, g(k,\eta)+ a^\dagger_{-\vk}\,  g^*(k,\eta)\,. \label{chik2} \ee With the product $\chi_{\vk}\,\chi_{-\vk}$, both the single-particle and correlated pair states in (\ref{rho2}) contribute to the power spectrum,  we find for de-Sitter with $C(\eta) = -1/H\eta$ and for super-Hubble modes
 \be \mathcal{P}_k(\eta) \simeq \Big(\frac{k^3\,H^2\,\eta^2}{2\pi^2} \Big)\,\frac{1}{2\,k^3\eta^2}\,\Big[1+ a\,Y^2\ln^2(-k\eta) + \cdots\Big] \simeq \frac{H^2}{4\pi^2} \,\Big[1+ a\,Y^2\ln^2(-k\eta) + \cdots\Big] \label{Pofkfina}\ee where $a$ is a constant that depends explicitly on the matrix elements. It will be obtained below from a more systematic treatment. This perturbative analysis, while preliminary and very approximate in the form of the secular logarithms,  yields a simple understanding of the physical processes that describe the reduced density matrix and the emergence of the entanglement entropy: the production of correlated fermion pairs kinematically entangled with inflaton fluctuations. It also  highlights in a simple but approximate manner the relationship between the corrections to the power spectrum and the entanglement entropy. The sections below provide a more technically   detailed derivation and confirmation  of these results from the effective action.

\section{Reduced density matrix}\label{sec:reduced}

The effective action for inflaton degrees of freedom obtained by tracing out fermionic degrees of freedom has been obtained in ref.\cite{boyfer}. For consistency and continuity in the presentation we summarize the main aspects of the derivation. The reader is referred to ref.\cite{boyfer} for more technical details.

 The time evolution of a   density matrix initially prepared at time $\eta_0$ is given by
 \be \rho(\eta)= U(\eta,\eta_0)\,\rho(\eta_0)\,U^{-1}(\eta,\eta_0) \,,\label{rhoeta}\ee where $\mathrm{Tr}[\rho(\eta_0)]=1$ and  $U(\eta,\eta_0)$ is the unitary time evolution operator of the full theory, it obeys
\be i\frac{d}{d\eta} U(\eta,\eta_0) = H(\eta) \,U(\eta,\eta_0)~~;~~ U(\eta_0,\eta_0) =1 \label{U} \ee where $H(\eta)$ is the total Hamiltonian. Therefore
\be U(\eta,\eta_0) = \mathbf{T}\Big[e^{-i\int_{\eta_0}^\eta H(\eta')d\eta'}\Big]~~;~~  U^{-1}(\eta,\eta_0) = \widetilde{\mathbf{T}}\Big[e^{i\int_{\eta_0}^\eta H(\eta')d\eta'}\Big] \ee with $\mathbf{T}$ the time-ordering symbol describing evolution forward in time and $\widetilde{\mathbf{T}}$ the anti-time ordered symbol describing evolution backwards in time.

Consider the initial density matrix at a conformal time $\eta_0$ and for the conformally rescaled fields
  to be of the form
\begin{equation}
 {\rho}(\eta_0) =  {\rho}_{\chi}(\eta_0) \otimes
 {\rho}_{\psi}(\eta_0) \,.\label{inidensmtx}
\end{equation} This choice while ubiquitous in the literature neglects possible initial correlations. We   consider an initial time $\eta_0$ such that physical wavelengths of cosmological relevance   were deep inside the Hubble radius at $\eta_0$. We will focus on the time evolution well after their physical wavelength have become super-Hubble during inflation when the amplitude of the scalar modes $\chi$  become amplified. Under this assumption initial correlations between these modes and the fermionic degrees of freedom are perturbatively small, hence we adopt this initially factorized density matrix with the understanding that the role of initial correlations between the inflaton and the fermionic degrees of freedom remains to be studied further.

  Since we are considering a de Sitter space-time, we take the initial time $\eta_0$ to be earlier than or equal to the time at which the slow-roll (nearly de Sitter) stage begins.

Our goal is to evolve this initial density matrix in (conformal) time obtaining (\ref{rhoeta}) and trace over the fermionic degrees of freedom  ($ \overline{\psi},\psi$) leading to a \emph{reduced} density matrix for   $\chi$ namely
\be \rho^r_\chi(\eta) = \mathrm{Tr}_{\psi} \rho(\eta)\,. \label{rhofired}\ee

There is no natural choice of the initial density matrices for the inflaton or fermionic fields, however,  consistently with  the analysis of the previous section and to  exhibit the main physical consequences of tracing over the fermionic degrees of freedom in the simplest setting we choose both fields to be in their respective Bunch-Davies vacuum state, namely
\be  {\rho}_{\chi}(\eta_0) = |0\rangle_{\chi}\,{}_\chi\langle 0|~~;~~  {\rho}_{\psi}(\eta_0) = |0\rangle_{\psi}\,{}_\psi\langle 0| \,.\label{BDinirho} \ee   This condition can be generalized straightforwardly.  In the discussion   below, we refer to $\psi,\overline{\psi}$ collectively as $\psi$ to simplify notation.

In the field basis the matrix elements of $ {\rho}_{\chi}(\eta_0);{\rho}_{\psi}(\eta_0)$
are given by
\begin{equation}
\langle \chi | {\rho}_{\chi}(\eta_0) | \widetilde{\chi}\rangle =
\rho_{\chi,0}(\chi ,\widetilde{\chi})~~;~~\langle \psi | {\rho}_{\psi}(\eta_0) | \widetilde{\psi}\rangle =
\rho_{\psi,0}(\psi ;\widetilde{\psi})\,, \label{fieldbasis}
\end{equation} and we have suppressed the coordinate arguments of the fields in the matrix elements. In this basis
\bea   \rho(\chi_f,\psi_f;\widetilde{\chi}_f,\widetilde{\psi}_f;\eta_f) & = &      \langle \chi_f;\psi_f|U(\eta_f,\eta_0) {\rho}(0)U^{-1}(\eta_f,\eta_0)|\widetilde{\chi}_f;\widetilde{\psi}_f\rangle \nonumber \\
& = & \int D\chi_i D\psi_i D\widetilde{\chi}_i D\widetilde{\psi}_i ~ \langle \chi_f;\psi_f|U(\eta_f,\eta_0)|\chi_i;\psi_i\rangle\,\rho_{\chi,0}(\chi_i;\widetilde{\chi}_i)\times \nonumber\\
&& \rho_{\psi,0}(\psi_i;\widetilde{\psi}_i)\,
 \langle \widetilde{\chi}_i;\widetilde{\psi}_i|U^{-1}(\eta_f,\eta_0)|\widetilde{\chi}_f;\widetilde{\psi}_f\rangle \,. \label{evolrhot}\eea The $\int D\chi$ etc, are functional integrals,  which for the fermionic degrees of freedom   are in terms of Grassmann valued fields and $D\psi \equiv D\psi D\overline{\psi}$. To simplify notation space-time arguments have been suppressed.   The matrix elements of the   forward and backward time evolution operators can be written as path integrals, namely
 \bea   \langle \chi_f;\psi_f|U(\eta_f,\eta_0)|\chi_i;\psi_i\rangle  & = &    \int \mathcal{D}\chi^+ \mathcal{D}\psi^+\, e^{i \int^{\eta_f}_{\eta_0} d\eta' d^3 x \mathcal{L}[\chi^+,\psi^+]}\,,\label{piforward}\\
 \langle  \widetilde{\chi}_i;\widetilde{\psi}_i|U^{-1}(\eta_f,\eta_0)|\widetilde{\chi}_f;\widetilde{\psi}_f\rangle &  =  &   \int \mathcal{D}\chi^- \mathcal{D}\psi^-\, e^{-i \int^{\eta_f}_{\eta_0}\int d^3 x \mathcal{L}[\chi^-,\psi^-]}\,,\label{piback}
 \eea where
 $ \mathcal{L}[\chi,\psi] $ can be read off (\ref{rescalagds})   and
 the boundary conditions on the path integrals are
  \bea     \chi^+(\vec{x},\eta_0)=\chi_i(\vec{x})~;~
 \chi^+(\vec{x},\eta_f)  &  =  &   \chi_f(\vec{x})\,,\nonumber \\   \psi^+(\vec{x},\eta_0)=\psi_i(\vec{x})~;~
 \psi^+(\vec{x},\eta_f) & = & \psi_f(\vec{x}) \,,\label{piforwardbc}\\
     \chi^-(\vec{x},\eta_0)= \widetilde{\chi}_i(\vec{x})~;~
 \chi^-(\vec{x},\eta_f) &  = &    \widetilde{\chi}_f(\vec{x})\,,\nonumber \\   \psi^-(\vec{x},\eta_0)=\widetilde{\psi}_i(\vec{x})~;~
 \psi^-(\vec{x},\eta_f) & = & \widetilde{\psi}_f(\vec{x}) \,.\label{pibackbc}
 \eea

 The fields $\chi^\pm,\psi^\pm$ describe the time evolution forward   ($+$) with $U(\eta,\eta_0)$  and backward  ($-$ ) with $U^{-1}(\eta,\eta_0)$,    this is the Schwinger-Keldysh formulation\cite{schwinger,keldysh,maha} of time evolution of density matrices.

 The reduced density matrix for   $\chi$ is obtained by tracing over fermionic degrees of freedom, namely
\be \rho^{r}(\chi_f,\widetilde{\chi}_f;\eta_f) = \int D\psi_f \,\rho(\chi_f,\psi_f;\widetilde{\chi}_f,\psi_f;\eta_f) \,.\label{rhored} \ee We find
\be \rho^{r}(\chi_f,\widetilde{\chi}_f;\eta_f) = \int D\chi_i   D\widetilde{\chi}_i  \,  \mathcal{T}[\chi_f,\widetilde{\chi}_f;\chi_i,\widetilde{\chi}_i;\eta_f;\eta_0] \,\rho_\chi(\chi_i,\widetilde{\chi}_i;\eta_0)\,,\label{rhochieta} \ee
where the time evolution kernel $\mathcal{T}$  is given by the following path integral representation
\be \mathcal{T}[\chi_f,\widetilde{\chi}_f;\chi_i,\widetilde{\chi}_i;\eta_f;\eta_0] = {\int} \mathcal{D}\chi^+ \,  \mathcal{D}\chi^- \, e^{i S_{eff}[\chi^+,\chi^-;\eta_f]}\,.\label{timevolredro} \ee  The total effective action that yields the time evolution of the reduced density matrix is given by
 \be S_{eff}[\chi^+,\chi^-;\eta_f] = \int^{\eta_f}_{\eta_0} d\eta' \int d^3x \Big[\mathcal{L}_0[\chi^+]- \mathcal{L}_0[\chi^-] \Big] + \mathcal{F}[\chi^+, \chi^-]\,, \label{Seff}\ee
The influence action $\mathcal{F}$ is defined by
 \be  e^{i\mathcal{F}[\chi^+;\chi^-]}   =   \int D\psi_i  \, D\widetilde{\psi}_i D\psi_f  \,\rho_{\psi}(\psi_i,\widetilde{\psi}_i;\eta_0) \, \int \mathcal{D}\psi^+ \mathcal{D}\psi^- \, e^{i   \int d^4x \Big\{\left[\mathcal{L}_+[\psi^+;\chi^+]-\mathcal{L}_-[\psi^-;\chi^-] \right] \Big\}}\,,\label{inffunc}\ee and we used the shorthand notation
 \be \mathcal{L}_{\pm}[\psi^\pm;\chi^\pm] = \mathcal{L}_0[\psi^\pm]-Y \chi^\pm (x) :\overline{\psi}^{\,\pm} (x)  \psi^\pm(x) : \,.\label{d4x} \ee The boundary conditions on the fermionic path integrals are
 \be \psi^+(\vec{x},\eta_0)=\psi_i(\vec{x})~;~
 \psi^+(\vec{x},\eta_f)=\psi_f(\vec{x})~~;~~ \psi^-(\vec{x},\eta_0)=\widetilde{\psi}_i(\vec{x})~;~
 \psi^-(\vec{x},\eta_f)=\widetilde{\psi}_f(\vec{x})={\psi}_f(\vec{x}) \,, \label{bcchis} \ee the last equality is a consequence of the trace.

The path integral  in the fermionic sector is  a representation of the time evolution forward and backwards of the fermionic density matrix, in (\ref{inffunc}) where $ \chi^\pm $ act  as   \emph{external sources} coupled to   $:\overline{\psi}^{\,\pm} (x)  \psi^\pm(x) :$. These sources   are different along the different branches,
\be e^{i\mathcal{F}[\chi^+;\chi^-]} = \mathrm{Tr}_{ \psi } \Big[ \mathcal{U}(\eta_f,\eta_0;\chi^+)\,\rho_\psi(\eta_0)\,  \mathcal{U}^{-1}(\eta_f,\eta_0;\chi^-) \Big]\,, \label{trasa}\ee where   $\mathcal{U}(\eta,\eta_0;\chi^\pm)$ is the   time evolution operator in the fermionic sector in presence of \emph{external sources} $\chi^\pm$ namely
 \be \mathcal{U}(\eta,\eta_0;\chi^+) = \mathbf{T}\Big( e^{-i \int_{\eta_0}^\eta H_\psi[\chi^+(\eta')]d\eta'}\Big) ~~;~~
\mathcal{U}^{-1}(\eta,\eta_0;\chi^-) = \widetilde{\mathbf{T}}\Big( e^{i \int_{\eta_0}^{\eta} H_\psi[\chi^-(\eta')]d\eta'}\Big)\,, \ee
and
\be H_\psi[\chi^\pm(\eta)] = H_{0 \psi}(\eta)+Y\,\int d^3x \, \chi^\pm(\vx,\eta)   :\overline{\psi}^{\,\pm} (\vx,\eta)  \psi^\pm(\vx,\eta) : \,.\label{timevchi}\ee  In eqn. (\ref{timevchi}) $H_{0\psi}(\eta)$ is the free field Hamiltonian for the field $\psi$ which depends explicity on time as a consequence of the $\eta$ dependent mass term  in the fermionic Lagrangian density    (\ref{l0psi}) and   $\chi^\pm$ in the interaction terms are \emph{classical} c-number sources.

 The calculation of $\mathcal{F}[\chi^+;\chi^-] $ proceeds  by passing to the interaction picture for the Hamiltonian $H_\psi[\chi^\pm(\eta)]$, defining
\be  \mathcal{U}(\eta;\eta_0;\chi^\pm) = \mathcal{U}_0(\eta;\eta_0) ~ \mathcal{U}_{ip}(\eta;\eta_0;\chi^\pm) \label{ipicture} \ee where $\mathcal{U}_0(\eta;\eta_0)$ is the time evolution operator of the free field $\psi$   and cancels out in the trace in (\ref{trasa}). The fermionic fields in $\mathcal{U}_{ip}(\eta;\eta_0;\chi^\pm)$ feature the free field time evolution (\ref{psiex}). The trace can be obtained systematically in perturbation theory in $Y$. Using the results of ref.\cite{boyeff,boyfer} we find up to $\mathcal{O}(Y^2)$  in the cumulant expansion

\bea i\mathcal{F}[\chi^+, \chi^-]  &  = &  -\,Y^2\int d^3x_1 d^3x_2 \int^{\eta_f}_{\eta_0} d\eta_1\,\int^{\eta_1}_{\eta_0} d\eta_2\,\Bigg\{ \chi^+(\vx_1,\eta_1)\chi^+(\vx_2,\eta_2)\,G^>(x_1;x_2)   \nonumber \\ & + &  \chi^-(\vx_1,\eta_1)\chi^-(\vx_2,\eta_2)\,G^<(x_1;x_2)  -   \chi^+(\vx_1,\eta_1)\chi^-(\vx_2,\eta_2)\,G^<(x_1;x_2)  \nonumber\\
  &- &   \chi^-(\vx_1,\eta_1)\chi^+(\vx_2,\eta_2)\,G^>(x_1;x_2)\Bigg\} ~~;~~ x_1 = (\eta_1,\vx_1) ~~ \mathrm{etc}\,,\label{Funravel}\eea where
  \begin{eqnarray}
&& {G}^>(x_1;x_2) =   \langle  \big(
: \overline{\psi}(x_1) \psi(x_1)::\overline{\psi}(x_2) \psi(x_2):\big)\rangle_{ \psi }     \,,\label{ggreat} \\&&  {G}^<(x_1;x_2) =   \langle \big(
: \overline{\psi}(x_2) \psi(x_2)::\overline{\psi}(x_1) \psi(x_1):\big)\rangle_{ \psi}    \,,\label{lesser} \eea

and the averages over fermionic variables are given by
\be  \langle
(\cdots) \rangle_{ \psi }  = \frac{\mathrm{Tr}_\psi(\cdots)\rho_\psi(\eta_0) }{\mathrm{Tr}_\psi \rho_\psi(\eta_0) }\,.\label{correzero}\ee We have used that normal ordering in the interaction picture yields
\be  \mathrm{Tr}_{\psi} (: \overline{\psi}(x) \psi(x): ) \rho_\psi(\eta_0) =0 \,,\label{noror} \ee
since the initial density matrix corresponds to the (Bunch-Davies) vacuum state for the fermionic degrees of freedom. Furthermore, comparing (\ref{ggreat}) and (\ref{lesser})  it follows that
\be G^>(x_1;x_2) = G^<(x_2;x_1)\,. \label{ident}\ee

 The fermionic correlation functions $G^{\lessgtr}$ are identified as the fermion loop that enters in the second order contribution of the perturbative density matrix (\ref{rho2}) (see fig.(\ref{fig:secondorder})), thereby establishing a direct relation between the perturbative approach of the previous section and the effective action.

   In a spatially flat FRW cosmology  spatial translational invariance implies that
  \be G^{\lessgtr}(x_1,x_2) = G^{\lessgtr}(\vx_1-\vx_2;\eta_1,\eta_2) \equiv \frac{1}{V} \sum_{\vp} \mathcal{K}^\lessgtr_{p}(\eta_1,\eta_2)\,e^{i\vp\cdot(\vx_1-\vx_2)} \,,\label{kernelsft}\ee
  it is straightforward to find that
  \be \mathcal{K}^<_p(\eta_1,\eta_2) = \Big(\mathcal{K}^>_p(\eta_1,\eta_2)\Big)^* \,. \label{relakapa}\ee
  Therefore we write the influence action in terms of spatial Fourier transforms in a volume $V$, with
  \be \chi^\pm(\vx,\eta)  \equiv \frac{1}{\sqrt{V}}\sum_{\vk} \chi^\pm_{\vk}(\eta) \,e^{-i\vk\cdot\vx} \, ,\label{fts}\ee and performing the spatial integrals we obtain
 \bea i\mathcal{F}[\chi^+, \chi^-]  &  = & -Y^2\,\sum_{\vk} \int^{\eta_f}_{\eta_0} d\eta_1\,\int^{\eta_1}_{\eta_0} d\eta_2\,\Bigg\{ \mathcal{K}^>_{k}(\eta_1;\eta_2) \Big[\chi^+_{\vk}(\eta_1)\chi^+_{-\vk}(\eta_2)- \chi^-_{\vk}(\eta_1)\chi^+_{-\vk}(\eta_2)\Big]\nonumber \\ & + & \mathcal{K}^<_{k}(\eta_1;\eta_2) \Big[\chi^-_{\vk}(\eta_1)\chi^-_{-\vk}(\eta_2)-\chi^+_{\vk}(\eta_1)\chi^-_{-\vk}(\eta_2) \Big]\Bigg\}\,.\label{Funfina}\eea In summary, the reduced density matrix for the inflaton field is given by
  \be \rho^{r}(\chi_f,\widetilde{\chi}_f;\eta_f) = \int D\chi_i   D\widetilde{\chi}_i  \,  {\int} \mathcal{D}\chi^+ \,  \mathcal{D}\chi^- \, e^{i S_{eff}[\chi^+,\chi^-;\eta_f]} \,\rho_\chi(\chi_i,\widetilde{\chi}_i;\eta_0)\,,\label{rhochietafina} \ee  where  the total effective action that yields the time evolution of the reduced density matrix is given by (\ref{Seff},\ref{Funfina}) and the boundary conditions on the path integrals are
   \bea     \chi^+(\vec{x},\eta_0)=\chi_i(\vec{x})~;~
 \chi^+(\vec{x},\eta_f)  &  =  &   \chi_f(\vec{x})\,,\nonumber \\
 \chi^-(\vec{x},\eta_0)    =      \widetilde{\chi}_i(\vec{x})~;~ \chi^-(\vec{x},\eta_f) &   =      & \widetilde{\chi}_f(\vec{x})\,.\label{chibc}
 \eea

 Although the reduced density matrix is obtained by tracing over the fermionic degrees of freedom, the total density matrix evolves in time via the \emph{unitary} time evolution operator, therefore \be \mathrm{Tr}\rho(\eta_f) = \mathrm{Tr}\rho(\eta_0)\,, \label{constrace}\ee where the \emph{total} trace corresponds to tracing over both  the fermionic and \emph{inflaton} degrees of freedom. Taking the initial density matrix to be given by equation  (\ref{inidensmtx}) with (\ref{BDinirho}) the relation (\ref{constrace}) yields
 \be \int D\chi_f \,  \rho^{r}(\chi_f,\widetilde{\chi}_f=\chi_f;\eta_f) = 1 \,.\label{unitarho}\ee

It is convenient to introduce the center of mass $\Psi_{\vk}(\eta)$ and relative ${R}_{\vk}(\eta)$ variables as
\be \Psi_{\vk}(\eta_1) = \frac{1}{2}\,(\chi^+_{\vk}(\eta_1)+\chi^-_{\vk}(\eta_1)) ~~;~~ {R}_{\vk}(\eta_1) = (\chi^+_{\vk}(\eta_1)-\chi^-_{\vk}(\eta_1))\,, \label{cmrelvars}\ee thus the path integral measure becomes $D\chi D\widetilde{\chi}= D\Psi DR$ and the  boundary conditions become
\bea  \Psi_{\vk}(\eta_0)     & \equiv &  \Psi_{\vk,i} = \frac{1}{2}\big(\chi_{\vk,i} + \widetilde{\chi}_{\vk,i}\big)~~;~~ \Psi_{\vk}(\eta_f)   \equiv   \Psi_{\vk,f}   = \frac{1}{2}\big(\chi_{\vk,f} +  \widetilde{\chi}_{\vk,f}\big) \label{psibc}\\
 R_{\vk}(\eta_0)   & \equiv &    R_{\vk,i}=  \big(\chi_{\vk,i} - \widetilde{\chi}_{\vk,i}\big)~~~;~~~ R_{\vk}(\eta_f)  \equiv R_{\vk,f}   = \big(\chi_{\vk,f} -  \widetilde{\chi}_{\vk,f}\big) \,.\label{Rbc} \eea

In terms of these variables   the effective action (\ref{Seff}) becomes
\bea && iS_{eff}[\Psi,R;\eta_f]   =   \sum_{\vk} \Bigg\{ \int^{\eta_f}_{\eta_0}\,d\eta_1 \,i\,\Big[R'_{\vk}(\eta_1)\, {\Psi'}_{-\vk}(\eta_1)- W^2(\eta)\,R_{\vk}(\eta_1)\,\Psi_{-\vk}(\eta_1)  \Big] - \nonumber \\  &&
\int^{\eta_f}_{\eta_0}\,d\eta_1 \int^{\eta_f}_{\eta_0}\,d\eta_2\,\Big[ \frac{1}{2} R_{\vk}(\eta_1)\,N_k(\eta_1;\eta_2)\,R_{-\vk}(\eta_2) + R_{\vk}(\eta_1)\,i \Sigma^R_k(\eta_1;\eta_2)\,\Psi_{-\vk}(\eta_2)    \Big]\Bigg\}\,,\label{iSeff2}\eea
where $' \equiv d/d\eta$ and
\bea W^2(\eta) & = & k^2 - \frac{1}{\eta^2}\big[\nu^2_\chi - \frac{1}{4}  \big] \,,\label{omega2} \\  {N}_k(\eta_1;\eta_2) & = &  \frac{{Y^2}}{2} \Big[\mathcal{K}^>_{k}(\eta_1;\eta_2)+ \mathcal{K}^<_{k}(\eta_1;\eta_2)\big] \label{Nker} \,,\\ \Sigma^R_k(\eta_1;\eta_2) & = &  \Sigma_k(\eta_1;\eta_2)\Theta(\eta_1-\eta_2)~;~ \Sigma_k(\eta_1;\eta_2)= {-iY^2}  \Big[\mathcal{K}^>_{k}(\eta_1;\eta_2)- \mathcal{K}^<_{k}(\eta_1;\eta_2)\big] \,. \label{sigmaker}\eea

The Gaussian path integrals over $\Psi, R$ are carried out by standard methods: introduce the classical paths $\Psi^c,R^c$ and fluctuations around them $z,r$ respectively as
\be \Psi_{\vk}(\eta_1) = \Psi^c_{\vk}(\eta_1)+z_{\vk}(\eta_1)~~;~~ R_{\vk}(\eta_1) = R^c_{\vk}(\eta_1)+r_{\vk}(\eta_1)\,,\label{classi}\ee where $\Psi^c_{\vk}(\eta_1); R^c_{\vk}(\eta_1)$ fulfill  the boundary conditions (\ref{psibc},\ref{Rbc}) and
\be z_{\vk}(\eta_0) = z_{\vk}(\eta_f)=0 ~~;~~ r_{\vk}(\eta_0) = r_{\vk}(\eta_f)=0 \,,\label{flucsbc}\ee and require that the \emph{linear} terms in $r_{\vk},z_{\vk}$ in $S_{eff}$ vanish. This yields the following equations of motion for $\Psi^c,R^c$
\be    \frac{d^2}{d\eta^2_1}\,\Psi^c_{\vk}(\eta_1)+W^2(\eta_1)\Psi^c_{\vk}(\eta_1)+\int^{\eta_1}_{\eta_0} \Sigma_k(\eta_1;\eta_2)\,\Psi^c_{\vk}(\eta_2)\,d\eta_2   =    i\,\int^{\eta_f}_{\eta_0} N_k(\eta_1;\eta_2)\,R^c_{\vk}(\eta_2)\,d\eta_2 \,,\label{Psieom}\ee and
\be  \frac{d^2}{d\eta^2_1}\,R^c_{\vk}(\eta_1)+W^2(\eta_1)R^c_{\vk}(\eta_1)+\int^{\eta_f}_{\eta_1} \Sigma_k(\eta_2;\eta_1)\,R^c_{\vk}(\eta_2)\,d\eta_2   =   0\,.\label{Reom}\ee Therefore the Gaussian path integrals yield
\be \int D\Psi DR\,e^{iS_{eff}[\Psi,R;\eta_f]} = \mathcal{N}(\eta_f)~~e^{iS_{eff}[\Psi_i,\Psi_f,R_i,R_f;\eta_f]}\,. \label{PItot}\ee The normalization factor
\be \mathcal{N}(\eta_f) = \int Dr Dz\,e^{iS_{eff}[z,r;\eta_f]} \label{norma}\ee only depends on $\eta_f$ but does \emph{not} depend on the initial and final values of the fields as a consequence of the boundary conditions (\ref{flucsbc}) for the fluctuations. This factor  does not need to be calculated because it is completely determined by the unitarity condition (\ref{unitarho}).

Using the equations of motion (\ref{Psieom},\ref{Reom}) we find
\bea iS_{eff}[\Psi_i,\Psi_f,R_i,R_f;\eta_f] & = &  \sum_{\vk}\Bigg\{  i\Big[R_{\vk,f}\,\Psi^{'c}_{-\vk}(\eta_f)-R_{\vk,i}\,\Psi^{'c}_{-\vk}(\eta_0)  \Big]\nonumber \\ & + & \frac{1}{2} \int^{\eta_f}_{\eta_0}d\eta_1 \int^{\eta_f}_{\eta_0}d\eta_2\, R^c_{\vk}(\eta_1)\,N_k(\eta_1,\eta_2)\,R^c_{\vk}(\eta_2) \Bigg\} \,, \label{Seffulti}\eea in this expression $\Psi^c,R^c$ are the solutions of the equations of motion (\ref{Psieom},\ref{Reom}) with the boundary conditions (\ref{psibc},\ref{Rbc}).

 To proceed further we need i) the initial density matrix $\langle \chi_i | {\rho}_{\chi}(\eta_0) | \widetilde{\chi}_i\rangle =
\rho_{\chi,0}(\chi_i ,\widetilde{\chi}_i)$  in the Schroedinger representation, ii) the kernels $ \mathcal{K}^\lessgtr_{k}(\eta_1,\eta_2)$, iii) the solution of the equations of motion (\ref{Psieom},\ref{Reom}) with the boundary conditions (\ref{psibc},\ref{Rbc}).

\subsection{Initial density matrix:}\label{sub:inirho}
From the expansion (\ref{chiex}), we define
\bea \chi_{\vk}(\eta) & = &  a_{\vk}\, g(k,\eta) + a^\dagger_{-\vk}\,g^*(k,\eta) \label{chik}\\
\chi'_{\vk}(\eta)  & = &  a_{\vk} \,g'(k,\eta) + a^\dagger_{-\vk}\,{g^*}'(k,\eta) \label{chikp}\eea where $g(k,\eta)$ obey the same wave equation as (\ref{chimodes}) and are given by (\ref{gqeta})   with Wronskian condition
\be g^*(k,\eta)\,g'(k,\eta) - {g^*}'(k,\eta)\,g (k,\eta) = -i \,. \label{wrons}\ee    The relations (\ref{chik},\ref{chikp}) can be inverted to yield
\bea a_{\vk} & = & i \Big[ g^*(k,\eta)\, \chi'_{\vk}(\eta) -  {g^*}'(k,\eta)\,\chi_{\vk}(\eta) \Big]\label{Opera} \\
 a^\dagger_{\vk} & = & -i \Big[ g(k,\eta)\, \chi'_{-\vk}(\eta) -  {g}'(k,\eta)\,\chi_{-\vk}(\eta) \,, \Big]\label{Operadag}   \eea since the operators $a_{\vk} \,, a^\dagger_{\vk}$ are independent of time the relation (\ref{Opera}) can be written at the initial time $\eta_0$ as
 \be a_{\vk}  =   i \Big[ g^*(k,\eta_0)\, \chi'_{\vk}(\eta_0) -  {g^*}'(k,\eta_0)\,\chi_{\vk}(\eta_0) \Big]\label{Operazero} \ee

 The Bunch-Davies vacuum obeys the condition (\ref{bdvac}) in the Schroedinger representation at the time $\eta_0$. The canonical momentum conjugate to $\chi$ is
 \be \pi_{ \vk} = \chi'_{ \vk} = \frac{\delta}{\delta \chi_{-\vk}} \,,\label{Schrep}\ee therefore the condition (\ref{bdvac}) becomes a functional differential equation for the vacuum Schroedinger wave-functional at $\eta_0$, $\Upsilon [\chi;\eta_0]= \langle \chi |0\rangle_{\chi}$, namely
 \be \Bigg[  \frac{\delta}{\delta \chi_{-\vk}} - i\Bigg( \frac{{g^*}'(k,\eta_0)}{g^*(k,\eta_0) }\Bigg)\,\chi_{\vk}(\eta_0) \Bigg] \Upsilon [\chi;\eta_0] = 0 \label{Scheqn} \ee with solution
 \be \Upsilon [\chi;\eta_0] =  {N}\,e^{-\frac{1}{2}\sum_{\vk}  \Omega_k ~\chi_{\vk} \, \chi_{-\vk} }\,, \label{wafu}\ee where
 \be \Omega_k =  -i \Big( \frac{{g^*}'(k,\eta_0)}{g^*(k,\eta_0) }\Big)\,,\label{omedef}\ee and
 $ {N}$ is a normalization factor. Therefore the initial density matrix for the $\chi$ field in the Schroedinger representation is given by
 \be \langle \chi_i | {\rho}_{\chi}(\eta_0) | \widetilde{\chi}_i\rangle =
\rho_{\chi,0}(\chi_i ,\widetilde{\chi}_i) = \Upsilon [\chi_i;\eta_0]\,\Upsilon^* [\widetilde{\chi_i};\eta_0] \,,\label{rhochini} \ee the normalization $\mathrm{Tr}\rho_\chi(\eta_0) =1$ fixes the value of $|N|$.

In terms of the center of mass and relative variables (see eqns. (\ref{psibc},\ref{Rbc}))
\be  \Psi_{\vk,i} = \frac{1}{2}\big(\chi_{\vk,i} + \widetilde{\chi}_{\vk,i}\big) ~~;~~  R_{\vk,i}=  \big(\chi_{\vk,i} - \widetilde{\chi}_{\vk,i}\big) \label{iniwig} \ee we find
\be \rho_{\chi,0}(\chi_i ,\widetilde{\chi}_i) \equiv \rho_{\chi,0}(\Psi_i ,R_i) = | {N}|^2\,\Pi_{\vk}~~e^{-\Omega_{R,k}\big[\Psi_{\vk,i}\Psi_{-\vk,i}+ \frac{1}{4}\, R_{\vk,i} R_{-\vk,i} \big] }~~ e^{-i\Omega_{I,k}\,\Psi_{\vk,i}R_{-\vk,i} }\,  \label{rhiniwig}\ee where $\Omega_R,\Omega_I$ are real and given by
\be \Omega_{R,k} = \frac{1}{2\,|g(k,\eta_0)|^2}~~;~~ \Omega_{I,k} = -\Omega_{R,k}\Big[{g^*}'(k,\eta_0) g(k,\eta_0)+ {g}'(k,\eta_0){g^*}(k,\eta_0)   \Big]\,.\label{omes}\ee

Finally from the expression (\ref{rhochietafina}), the reduced density matrix in terms of the center of mass and relative variables is given by
\be \rho^r(\Psi_f,R_f;\eta_f) = \mathcal{N}(\eta_f)\, \int D\Psi_i\,DR_i ~ e^{iS_{eff}[\Psi_i,\Psi_f,R_i,R_f;\eta_f]}~\rho_{\chi,0}(\Psi_i ,R_i)\,.\label{rhofinawig}\ee This is the final form of the reduced density matrix, where the functional integrals are simple Gaussian integrals that can be   carried out once the fermionic correlation functions and the solutions of the equations of motion are obtained. The normalization pre-factor in (\ref{rhofinawig}) is determined from the condition of unitary time evolution (\ref{unitarho}), namely
\be \int D\Psi_f\,\rho^r(\Psi_f,R_f=0;\eta_f) = 1 \,.\label{unitawig}\ee

\subsection{Fermionic correlations:}\label{sub:fermi}
The kernels $\mathcal{K}^{\lessgtr}_k$ defined as the spatial Fourier transforms of the Fermionic correlation functions (\ref{kernelsft}) are obtained from the mode functions $U,V$ in the field expansion (\ref{psiex}), given by (\ref{Us},\ref{Vs},\ref{Up}) with (\ref{fketasolu}). For generic fermion mass $m_f$ the kernels do not feature a useful analytic expression, however assuming that the inflation scale is much larger than the typical mass scales of the standard model (and even beyond), we focus on the case $m_f\ll H$. In this case we use the results of ref.\cite{boyfer} and $\Sigma_k,N_k$ in (\ref{Nker}, \ref{sigmaker}) are given by
 \be  {N}_k(\eta_1;\eta_2)=  \frac{Y^2}{8\pi}\Big[\frac{d^2}{d\eta_1\,d\eta_2} - k^2 \Big]\Bigg\{ \delta(\eta_1-\eta_2)- \frac{1}{\pi} \frac{\sin[k(\eta_1-\eta_2)]}{(\eta_1-\eta_2)}\Bigg \} \label{noisiker}\ee
\be \Sigma_k(\eta_1;\eta_2) = \frac{Y^2}{8\pi^2}\Big[\frac{d^2}{d\eta_1\,d\eta_2} - k^2 \Big]\Bigg \{\cos[k(\eta_1-\eta_2)]\, \frac{d}{d\eta_2} \ln\Big[\frac{(\eta_1-\eta_2)^2+\epsilon^2}{(-\eta_0)^2} \Big]
\Bigg\}\,, \label{sigretker}\ee where $\epsilon\rightarrow 0^+$ is a short-distance regulator and $-\eta_0$ is a renormalization scale chosen to coincide with the initial time.

\subsection{Solutions of the equations of motion:}
We solve the equations of motion (\ref{Psieom},\ref{Reom}) in a perturbative expansion in $Y^2$ and insert these solutions in (\ref{Seffulti}) to obtain the effective action up to order $Y^2$. We begin with the zeroth-order solution to highlight several relevant aspects and shed light on the interpretation in the interacting case.

\vspace{1mm}

\textbf{Zeroth order solutions:}

The solutions of the zeroth-order equations of motion correspond to setting $Y^2=0$, namely $\Sigma_k,N_k =0$ in (\ref{Psieom},\ref{Reom}), yielding    the free field equation of motion (\ref{chimodes}) whose solutions are the mode functions $g(k,\eta)$ given by (\ref{gqeta}).

However, instead of using these complex mode functions, and  in order to separate the real from the purely imaginary contribution to the effective action we use the real mode functions
\be g_+(k,\eta) = \Big[-\frac{\pi\eta}{2}\Big]^{1/2} \,Y_{\nu_\chi}(-k\eta)~~;~~ g_-(k,\eta) = \Big[-\frac{\pi\eta}{2}\Big]^{1/2} \,J_{\nu_\chi}(-k\eta) \,,\label{gpm}\ee which describe the growing ($g_+$) and decaying ($g_-$) solutions for super-Hubble modes and satisfy  the Wronskian condition
 \be g^{'}_+(k,\eta)\,g_-(k,\eta)-g^{'}_-(k,\eta)\,g_+(k,\eta) = -1\,. \label{Wronk}\ee These real mode functions are related to the complex mode functions $g(k,\eta)$ (\ref{gqeta}) as
 \be g(k,\eta) = \frac{i}{\sqrt{2}}\,e^{i\frac{\pi}{2}(\nu_\chi+1/2)}\,\Big[g_+(k,\eta)-i g_-(k,\eta) \Big] \,.  \label{relages}\ee

  For a (nearly) massless inflaton field for which $\nu_\chi = 3/2$, in the super-Hubble limit $-k\eta \rightarrow 0^+$ these solutions behave as
\be g_+(k,\eta) =  \frac{1}{k^{3/2}\,\eta} ~~;~~ g_-(k,\eta) =  \frac{1}{3}\,k^{3/2}\,\eta^2 \,.\label{gpmSH}\ee In terms of these   mode functions the general solution of (\ref{Psieom}) for $Y^2=0$   is given by
\be \Psi^c_{\vk}(\eta_1) = Q_k \, g_+(k,\eta_1)+P_k \, g_-(k,\eta_1)\,, \label{psisolzero}\ee with the coefficients $Q_k,P_k$ fixed by the boundary conditions (\ref{psibc}). We find
\be \Psi^c_k(\eta_1) = \Psi_{\vk,i} \,  \frac{D_k[\eta_f;\eta_1]}{D_k[\eta_f;\eta_0]}   + \Psi_{\vk,f}\,   \frac{D_k[\eta_1;\eta_0]}{D_k[\eta_f;\eta_0]}   \,,\label{solpsizero} \ee where we introduced
\be D_k[\eta_1;\eta_2] = g_+(k,\eta_1)\,g_-(k;\eta_2)-g_+(k;\eta_2)\,g_-(k,\eta_1)\,.  \label{Deta}\ee The equation of motion for $R^c$ for $Y^2 =0$ with the boundary conditions (\ref{Rbc}) has a similar solution,
\be R^c_k(\eta_1) = R_{\vk,i} \,  \frac{D_k[\eta_f;\eta_1]}{D_k[\eta_f;\eta_0]}  + R_{\vk,f}\,  \frac{D_k[\eta_1;\eta_0]}{D_k[\eta_f;\eta_0]}   \,.\label{solRzero} \ee

Our goal is to obtain the effective action and entanglement entropy to leading order in $Y^2$, therefore we input the zeroth-order solution (\ref{solRzero}) for $R^c$ in the second line in (\ref{Seffulti}) because $N_k \propto Y^2$.

\vspace{1mm}

\textbf{Perturbative solution:}

As discussed in detail in ref.\cite{boyfer} upon integration by parts the self-energy term in
the equation of motion (\ref{Psieom}) becomes
\bea && \int_{\eta_0}^{\eta_1}   \Sigma_k(\eta_1;\eta_2)\Psi^c_{\vec k}(\eta_2) \,d\eta_2   =   -\frac{Y^2}{4\pi^2}  \frac{\Psi^c_{\vec k}(\eta_1)}{\epsilon^2} + \frac{Y^2}{4\pi^2}\ln\Big[\frac{(-\eta_0)}{\epsilon}\Big] \Bigg[ \frac{d^2 \Psi^c_{\vec k}(\eta_1)}{d\eta^2_1 }+k^2 \Psi^c_{\vec k}(\eta_1)\Bigg] \nonumber \\ &   & ~~~~~~~~~~+ \frac{Y^2}{4\pi^2} \int^{\eta_1}_{\eta_0}  \,\ln\Big[\frac{\eta_1-\eta_2}{(-\eta_0)} \Big]\, \frac{d}{d\eta_2} \Bigg\{ \cos[k(\eta_1-\eta_2)]\Bigg[\frac{d^2\Psi^c_{\vec k}(\eta_2)}{d\eta^2_2}+ k^2 \Psi^c_{\vec k}(\eta_2)\Bigg]\Bigg\}\,d\eta_2\,. \label{selfiterm} \eea In obtaining this expression, we have neglected the contribution from the lower limit ($\eta_0$) in the integration by parts, these contributions are finite and perturbatively small (since the mode functions are assumed to be deeply sub-Hubble at the initial time) as $\eta_1\rightarrow 0$ which is the limit of interest in this work. As discussed in ref.\cite{boyfer} the first two terms are absorbed into mass and wavefunction renormalization.


In particular, as shown in this reference, after absorbing the quadratic divergence  $\propto 1/\epsilon^2$ independent of $\eta_0$ into an intermediate renormalized mass $\widetilde{M}^2$ the fully renormalized mass (up to one loop) $M_R(\eta_0)$ obeys the relation
\be \frac{\widetilde{M}^2}{H^2} = \frac{M^2_R(\eta_0)}{H^2} - \frac{Y^2}{2\pi^2}\,\ln\Bigg[\frac{(-\eta_0)}{\epsilon} \Bigg]\,. \label{Mren}\ee Because $\widetilde{M}^2$ \emph{does not} depend on $\eta_0$, the combination on the right hand side of (\ref{Mren}) is \emph{invariant under a change of scale} $\eta_0$. As discussed in ref.\cite{boyfer}, the scale $\eta_0$ is chosen so that the renormalized mass $M^2_R(\eta_0) =0$, and choosing $-\eta_0$ to coincide with the onset of slow roll inflation yields a power spectrum that is scale invariant for $Y=0$ and the departure from scale invariance is a consequence of the interaction. See discussion in section (\ref{sec:discussion}) below.


Since we are primarily concerned with the asymptotic super-Hubble limit, we adopt here the renormalization procedure detailed in ref.\cite{boyfer} absorbing these  two terms  in the corresponding renormalizations and focus solely on the contribution from the third term in (\ref{selfiterm}).

We consider a perturbative solution of (\ref{Psieom}) of the form,
\be \Psi^c_{\vk}(\eta_1) = \Psi^c_{\vk,0}(\eta_1) +  \, \Psi^c_{\vk,1}(\eta_1) + \cdots
\label{pertsolPsi}\ee where $ \Psi^c_{\vk,0}(\eta_1)$ is given by the zeroth-order solution (\ref{solpsizero}), $\Psi^c_{\vk,1}(\eta_1) \propto Y^2$ etc. After renormalization (see details in ref.\cite{boyfer}) the first order correction  $\Psi^c_{\vk,1}$ obeys the equation
\be    \frac{d^2}{d\eta^2_1}\,\Psi^c_{\vk,1}(\eta_1)+W^2(\eta_1)\Psi^c_{\vk,1}(\eta_1)= I[k;\eta_1]   +   i\,R_{\vk,i} \, \xi_{k,1}(\eta_1) + i\,R_{\vk,f} \, \xi_{k,2}(\eta_1)\,,\label{Psi1eom}\ee where
\be I[k;\eta_1] = -\frac{Y^2}{2\pi^2}\int^{\eta_1}_{\eta_0} d\eta_2 \, \ln\Big[\frac{(\eta_1-\eta_2) }{(-\eta_0)}\Big]\,\frac{d}{d\eta_2} \Bigg\{\cos[k(\eta_1-\eta_2)]\frac{\Psi^c_{\vk,0}(\eta_2)}{\eta^2_2}  \Bigg\} \,,\label{inhohiera1} \ee and we have used the zeroth order equation of motion (\ref{Psieom}) (with $\Sigma_k=0; N_k =0$)  with $W^2_k(\eta_1) = k^2 -2/\eta^2_1$ neglecting $M^2_R/H^2 \ll 1$ with $M^2_R$ the renormalized inflaton mass, and
\bea \xi_{k,1}(\eta_1) &  =  & \int^{\eta_f}_{\eta_0}\,N_k(\eta_1;\eta_2)\, \frac{D_k[\eta_f ;\eta_2]}{D_k[\eta_f ;\eta_0]} \,d\eta_2 \label{xi1}\\
\xi_{k,2}(\eta_1) &  =  & \int^{\eta_f}_{\eta_0}\,N_k(\eta_1;\eta_2)\, \frac{D_k[\eta_2 ;\eta_0]}{D_k[\eta_f ;\eta_0]} \,d\eta_2 \label{xi2}\,. \eea

For the growing mode $\Psi^c_{\vk,0}(\eta_2) = g_+(k,\eta_2)$ we find in the super-Hubble limit $-k\eta_1 \rightarrow 0^+$
 \be I_+[k;\eta_1] = -\frac{Y^2}{2\pi^2}\,\frac{g_+(k,\eta_1)}{ \eta^2_1} \Bigg( \ln\Big[\frac{\eta_1}{\eta_0} \Big]-\frac{3}{2}\Bigg)+\cdots\,, \label{inho1stor} \ee  where the dots stand for subleading terms in this limit, whereas for the decaying mode $\Psi^c_{\vk,0}(\eta_2) = g_-(k,\eta_2)$   we find that $ I_-[k;\eta_1] \rightarrow Y^2 \times (\mathrm{constant}) $ in the same limit, in other words without secular terms.

The inhomogeneous equation (\ref{inhohiera1}) can be solved by introducing the   Green's function of the differential operator on the left hand side of (\ref{inhohiera1}) with retarded boundary conditions,
\be G^R_k(\eta_1,\eta_2) = G_k[\eta_1,\eta_2]\,\Theta(\eta_1-\eta_2)\,, \label{Gret}\ee where
\be G_k[\eta_1,\eta_2] = -D_k[\eta_1,\eta_2]\,. \label{GeqD}\ee
In the super-Hubble limit of both arguments we find
\be G_k[\eta_1,\eta_2] \rightarrow   \frac{1}{3} \Big[\frac{\eta^2_1}{\eta_2}-\frac{\eta^2_2}{\eta_1} \Big]   \,.  \label{greenfun}\ee In terms of this Green's function it follows that
\be \Psi^c_{\vk,1}(\eta_1)= \int^{\eta_1}_{\eta_0} G_k[\eta_1,\eta_2] \Bigg\{I[k;\eta_2]   +   i\,R_{\vk,i}\,  \xi_{k,1}(\eta_2) + i\,R_{\vk,f} \, \xi_{k,2}(\eta_2)\Bigg\} \,d\eta_2~~;~~ \Psi^c_{\vk,1}(\eta_0)=\Psi^{\,'\,c}_{\vk,1}(\eta )\Big|_{\eta_0}=0 \,.  \label{psifo}\ee

 Writing the zeroth order solution as a combination of $g_{\pm}$ it is straightforward to find that up to and including $\mathcal{O}(Y^2)$
\be \Psi^c_{\vk}(\eta_1) = Q_k \,\widetilde{g}_+(k,\eta_1)+ P_k \,\widetilde{g}_-(k,\eta_1)+ i \, R_{\vk,i}\,h_1(k,\eta_1)+i\,  R_{\vk,f}\,h_2(k,\eta_1) \,, \label{psitota}\ee where $Q_k, P_k$ are coefficients fixed by the boundary conditions (\ref{psibc}) and the $\widetilde{g}_\pm$ are the perturbatively corrected mode functions with the following limits
\bea  \widetilde{g}_\pm(k,\eta_0)  & = & g_\pm(k,\eta_0) \label{tilgszero}\\
\widetilde{g}_\pm(k,\eta_f) & =  & g_\pm(k,\eta_f)\,\Big[ 1 + Y^2 \,\mathcal{F}_\pm(k,\eta_f)+\cdots \Big] \,.\label{tilgsfin}\eea Using the results of ref.  \cite{boyfer} we find that  in the super-Hubble limit $-k\eta_f \rightarrow 0^+$
\bea \mathcal{F}_+(k,\eta_f) & = & \frac{1}{12\pi^2}\Big\{ \ln^2\big(-k\eta_f \big)-2 \ln\big( -k\eta_f\big)\ln\big(-k\eta_0 \big) \Big\}+\cdots\label{Fplus} \\
\mathcal{F}_-(k,\eta_f) & = & Y^2 \times (\mathrm{finite}\,\,\mathrm{constant})\,, \label{Fmin}\eea the dots in (\ref{Fplus}) stand for subleading terms in the super-Hubble limit. The functions $h_{1,2}(k,\eta_1)$ in eqn. (\ref{psitota}) are obtained from the integrals in (\ref{psifo}) and satisfy
\be h_{1,2}(k,\eta_0) = 0 ~~;~~ h'_{1,2}(k,\eta)\Big|_{\eta_0}=0 \,, \label{inihs}\ee their explicit expressions are given in   appendix (\ref{app:coeffs})    (see \ref{h1},\ref{h2}). Fixing the coefficients $Q_k, P_k$ to satisfy the boundary conditions (\ref{psibc}) yield
\be \Psi^c_{\vk}(\eta_1) = \Psi_{\vk,i} \, \frac{\widetilde{D}_k[\eta_f;\eta_1]}{\widetilde{D}_k[\eta_f;\eta_0]}   + \Psi_{\vk,f}\,  \frac{\widetilde{D}_k[\eta_1;\eta_0]}{\widetilde{D}_k[\eta_f;\eta_0]}  + i \, R_{\vk,i}\,H_1(k,\eta_1)+i\,  R_{\vk,f}\,H_2(k,\eta_1) \,,\label{psicfina}\ee with
\be \widetilde{D}_k[\eta_1;\eta_2] = \widetilde{g}_+(k,\eta_1)\,\widetilde{g}_-(k;\eta_2)-\widetilde{g}_+(k;\eta_2)\,\widetilde{g}_-(k,\eta_1)\,, \label{tilDeta}\ee and\footnote{Since $h_{1,2} \propto Y^2$ we considered the mode functions to zeroth order.}
\be H_{1,2}(k,\eta_1) = h_{1,2}(k,\eta_1) - h_{1,2}(k,\eta_f)\,  \frac{ {D}_k[\eta_1;\eta_0]}{ {D}_k[\eta_f;\eta_0]}  ~~;~~ H_{1,2}(k,\eta_0) = H_{1,2}(k,\eta_f)=0 \,.  \label{Hfinas}\ee

Gathering the results above, the effective action (\ref{Seffulti}) becomes
\bea iS_{eff}[\Psi_i,\Psi_f,R_i,R_f;\eta_f] & = &  \sum_{\vk}\Bigg\{  i R_{\vk,f}\,\Big( \Psi_{-\vk,i}\,A_{k,f}+ \Psi_{-\vk,f} \,B_{k,f} + i R_{\vk,i}\,C_{k,f} + i R_{\vk,f}\, D_{k,f} \Big)\nonumber \\ & - &  i R_{\vk,i}\,\Big( \Psi_{-\vk,i}\,A_{k,i}+ \Psi_{-\vk,f} \,B_{k,i} + i R_{\vk,i}\,C_{k,i} + i R_{\vk,f}\, D_{k,i} \Big) \nonumber \\ & + & \frac{1}{2} \, R^2_{\vk,i}\, J_{1,k} + \frac{1}{2} \, R^2_{\vk,f}\, J_{2,k} +  R_{\vk,i} \, R_{\vk,f}\, J_{3,k} \Bigg\} \,, \label{Seffulti2}\eea where the explicit form of the various coefficients is given in appendix (\ref{app:coeffs}). An important aspect of these coefficients is that they are all \emph{real}, this is the main advantage of having introduced the real mode functions $g_\pm(k,\eta)$.

The Gaussian functional integrals in (\ref{rhofinawig}) can now be carried out, yielding the   reduced density matrix
\be \rho^r(\Psi_f,R_f;\eta_f) = \widetilde{\mathcal{N}}\,\, \Pi_{\vk} \, \,\exp\Big\{-\alpha_k\,\Psi_{\vk,f}\Psi_{-\vk,f} - {\beta_k} \, R_{\vk,f}R_{-\vk,f}   - i \gamma_k \, \Psi_{\vk,f}R_{-\vk,f}\Big\} \label{rhorfi}\ee with $\widetilde{\mathcal{N}}$ a normalization factor determined by the condition (\ref{unitawig}),  and
\bea \alpha_k  & = &  \frac{B^2_{k,i}}{4\omega_k} \label{alfak}\\
\beta_k & = & \frac{1}{4\omega_k}\Bigg[J_{3,k}+D_{k,i}-C_{k,f} + A_{k,f}\,\frac{(A_{k,i}+\Omega_{I,k})}{2\Omega_{R,k}} \Bigg]^2 + \frac{J_{2,k}}{2}-D_{k,f} - \frac{A^2_{k,f}}{4\Omega_{R,k}} \label{betak} \\
\omega_k & = & \frac{\Omega_{R,k}}{4} + \frac{(A_{k,i}+\Omega_{I,k})^2}{4\Omega_{R,k}}-C_{k,i}-\frac{J_{1,k}}{2} \,. \label{omek} \eea Where $\Omega_{R,k},\Omega_{I,k}$ are given in terms of the real mode functions at $\eta_0$ by eqns.  (\ref{OmegaRzero},\ref{ratiozero}) in appendix (\ref{app:coeffs}) respectively. We do not quote the expression for $\gamma_k$ which is cumbersome and, as explained   in detail below, not relevant either for the power spectrum or the entanglement entropy.

The coefficients $J,D, C $ are all of $\mathcal{O}(Y^2)$ and $A, B \simeq \mathcal{O}(1) + \mathcal{O}(Y^2)$, therefore the squared term in $\beta_k$ must be computed up to $\mathcal{O}(Y^2)$.

\vspace{1mm}

\subsection{$\mathbf{ Y=0 }$:}

Before we analyze the reduced density matrix including the contribution from the fermionic correlations, it will prove informative to consider first the $Y=0$ case for which $J_{1,2,3}= C_{k,i}= C_{k,f}=D_{k,i}=D_{k,f}=0$. The coefficients $A, B$ for this case are gathered in appendix (\ref{app:yzero}) (see eqns. (\ref{Aiyzero}-\ref{dfyzero})), and the coefficient $\gamma_k$ in (\ref{rhorfi}) is given by
\be \gamma_k = \frac{1}{\Omega_{R,k}D^2[\eta_f,\eta_0]}\,\Big[A_{k,i}+\Omega_{I,k}+\Omega_{R,k}\,D[\eta_f,\eta_0] \Big]\,. \label{gamak}\ee Using the results for the coefficients in appendix (\ref{app:yzero}) and after straightforward algebra, we find
\bea \alpha_k & = & \Omega_{R,k}(\eta_f) \label{alfayzero}\\
\beta_k & = & \frac{\alpha_k}{4}   \label{betayzero}\\
\gamma_k & = & \Omega_{I,k}(\eta_f)\,,\label{gamayzero}\eea  where $\Omega_{R,k}(\eta_f),\Omega_{I,k}(\eta_f)$ are the coefficients (\ref{OmegaRzero},\ref{ratiozero}) with $\eta_0 \rightarrow \eta_f$. In terms of the complex mode functions $g(k,\eta)$ (\ref{gqeta}) related to $g_\pm(k,\eta)$ by eqn. (\ref{relages}), these coefficients are the same as those of eqn. (\ref{omes}) with $\eta_0\rightarrow \eta_f$. Replacing these coefficients in the reduced density matrix (\ref{rhorfi}) we find that it has \emph{exactly} the same form as the \emph{initial} density matrix (\ref{iniwig}) but with $\Psi_{\vk,i}, R_{\vk,i}  \rightarrow \Psi_{\vk,f}, R_{\vk,f}$ and $\Omega_{R,k},\Omega_{I,k} \rightarrow \Omega_{R,k}(\eta_f),\Omega_{I,k}(\eta_f)$. In other words, the reduced density matrix for $Y=0$ at $\eta = \eta_f$  is simply the initial pure state density matrix evolved in time from $\eta_0$ up to $\eta_f$  with the free field Hamiltonian,  namely
\be \rho^r(\chi_f,\widetilde{\chi}_f;\eta_f)
  = \Upsilon [\chi_f;\eta_f]\,\Upsilon^* [\widetilde{\chi_f};\eta_f] \,,\label{rhochfina} \ee where $\Upsilon$ is the Schroedinger wavefunctional describing the Bunch-Davies vacuum state at time $\eta_f$.

  Of course this is expected, in absence of interactions the reduced density matrix is simply the initial density matrix propagated in time with the unitary time evolution operator. However, it is reassuring, as well as an important check,  that the formalism described above yields the expected result in the non-interacting limit.

\subsection{$\mathbf{ Y\neq 0 }$}

The term proportional to $\gamma_k$ (a \emph{real} coefficient) in the exponent in (\ref{rhorfi}) is purely imaginary corresponding to a pure phase in the final density matrix which does not contribute to the power spectrum or the entanglement entropy and will be neglected in the analysis below. While this final expression for the reduced density matrix (\ref{rhorfi}) with the coefficients (\ref{alfak},\ref{betak},\ref{omek}) is cumbersome and unwieldy, we are primarily focused on the super-Hubble limit, where progress can be made by analyzing the behavior of the various coefficients to extract the leading behavior. The details of such analysis are provided in appendix (\ref{app:shcoeffs}) with the main result to leading order given by (see the final equations (\ref{alfakfinal},\ref{betakfinal}))
\bea   \alpha_k  & = &   = \frac{1}{\,\widetilde{g}^{\,2}_+(k,\eta_f)} \label{alfakrhor}\\
\beta_k & = & \frac{\alpha_k}{4}\,\Big[1+Y^2\,\mathcal{F}_+(k,\eta_f)\Big]\,. \label{betakrhor}\eea

\section{Entanglement entropy from the effective action: }

Going back to the original variables $\chi_f,\widetilde{\chi}_f$ (see equations (\ref{psibc},\ref{Rbc})) the final reduced density matrix reads \bea \rho^r(\chi_f,\widetilde{\chi}_f;\eta_f)& = & \widetilde{N}~ \Pi_k \exp\Bigg\{-\Big[\Big(\frac{\alpha_k}{4}+\beta_k \Big)\, \Big(\chi_{\vk,f}\,\chi_{-\vk,f}+\widetilde{\chi}_{\vk,f}\,\widetilde{\chi}_{-\vk,f} \Big)-2\,\Big(\beta_k-\frac{\alpha_k}{4} \Big)\,\chi_{\vk,f}\,\widetilde{\chi}_{-\vk,f}   \Big]  \Bigg\}\nonumber \\ & \times &  ~\exp\Big\{-i\gamma_k\,\Big(\chi_{\vk,f}\,\chi_{-\vk,f}-\widetilde{\chi}_{\vk,f}\,\widetilde{\chi}_{-\vk,f} \Big)  \Big\}\,. \label{rhorchifi} \eea This expression makes manifest that if $\beta_k = \alpha_k/4$ the reduced density matrix describes a \emph{pure} state since it is identified as the product of a wavefunctional times its complex conjugate. Therefore the results (\ref{alfakrhor},\ref{betakrhor}) imply that in presence of interactions the reduced density matrix describes a \emph{mixed} state. This is in agreement with the perturbative calculation in section (\ref{sec:pert}), and the discussion for the $Y=0$ case above.

The entanglement entropy of this mixed state is the Von-Neumann  entropy, it is given by
\be S_{vN} = -\sum_n \, \lambda_n \,\ln(\lambda_n) \,,\label{SVN} \ee where $\lambda_n$ are the eigenvalues of the density matrix, namely
\be \int D\widetilde{\chi}_f \,\rho^r(\chi_f,\widetilde{\chi}_f;\eta_f) \,\Phi_n(\widetilde{\chi}_f) \,= \lambda_n\, \Phi_n({\chi}_f) \,, \label{eigen}\ee the normalization condition (\ref{unitarho}) yields $\sum_{n} \lambda_n = 1$.   The second line in (\ref{rhorchifi}), i.e.  the phase, does not contribute to the eigenvalue equation  because it can be absorbed into the wavefunctions,  $ e^{i\gamma_k \chi\,\chi} \, \Phi_n( {\chi}_f) \rightarrow \Phi_n( {\chi}_f)$. Therefore this total phase can be safely set to zero in the reduced density matrix, it does not contribute to either the entropy or the power spectrum.
The entropy of gaussian density matrices has been originally obtained in the seminal work of refs.\cite{bombelli,srednicki,callan,cardy,neu}. We present an alternative to these methods that allows to establish a closer relation to mixed states in statistical physics. We recognize that setting to zero the phase in (\ref{rhorchifi})and introducing the definitions
\be \frac{\alpha_k}{4}+\beta_k  \equiv \frac{\mathcal{W}_k(\eta_f)}{2}\,\coth\Big[ \frac{\mathcal{W}_k(\eta_f)}{T_k(\eta_f)}\Big]~~;~~
\beta_k-\frac{\alpha_k}{4} \equiv \frac{\mathcal{W}_k(\eta_f)}{2\sinh\Big[ \frac{\mathcal{W}_k}{T_k(\eta_f)}\Big]} \,, \label{WTdefs} \ee yielding
\be \mathcal{W}_k(\eta_f) = 2 \Big[ \beta_k\,\alpha_k\Big]^{1/2} ~~;~~ \exp\Bigg[-\frac{\mathcal{W}_k(\eta_f)}{T_k(\eta_f)} \Bigg] = \Bigg[\frac{\Big(\frac{4\beta_k}{\alpha_k}\Big)^{1/2}-1}{\Big(\frac{4\beta_k}{\alpha_k}\Big)^{1/2}+1} \Bigg] \equiv \xi_k \,,  \label{WandT} \ee the reduced density matrix (\ref{rhorchifi})
is similar in form  to the Schroedinger representation of \emph{the density matrix of decoupled harmonic oscillators, each  in thermal equilibrium} with temperature $T_k(\eta_f)$\cite{feynman}, namely
\be \rho^r(\chi_f,\widetilde{\chi}_f;\eta_f) = Z^{-1}~\langle \chi_f \Big|\, \exp\Big\{-\sum_{\vk}\, \frac{H_k(\eta_f)}{T_k(\eta_f)}\Big\}\, \Big| \widetilde{\chi}_f\rangle \,,\label{rhohos}\ee with
\be H_k(\eta_f) = \frac{1}{2} \Big[ \,\pi_{\vk} \pi_{-\vk}+ \,\mathcal{W}_k(\eta_f)\,\chi_{\vk}\,\chi_{-\vk}\Big]\,, \label{Hchi}\ee $Z^{-1}$ is the normalization factor, and $\pi_{-\vk}$ is the canonical momentum conjugate to $\chi_{\vk}$.  As $Y^2 \rightarrow 0$ it follows from equation (\ref{betakrhor}) that  $\beta_k \rightarrow   \alpha_k/4$ and $T_k(\eta_f) \rightarrow 0$, therefore we  recover  the ground state density matrix as discussed above.

  The     eigenfunctions up to a normalization factor are
\be \Phi_n(\chi_f) \propto H_n\big[ \sqrt{\mathcal{W}_k} \,\chi_f\big]\,\exp[-\mathcal{W}_k\,(\chi_f)^2 ] \,, \label{WF}\ee where $H_n$ are Hermite polynomials, with eigenvalues
\be \lambda_n = \big[ 1-\xi_k \big]\, \xi^n_k \,,\label{pns}\ee  where $\xi_k$ is given by eqn. (\ref{WandT}) and we used the normalization condition (\ref{unitarho}). This normalization condition along with the expression for the entanglement entropy imply that the eigenvalues must fulfill the conditions $0 \leq \lambda_n < 1$. The fulfillment of this condition is discussed in detail in section (\ref{sec:discussion}) below.

 These results agree with those of refs.\cite{bombelli,srednicki,callan} obtained with different methods.  Finally the entanglement or Von-Neumann entropy is given by
\be S_{vN} = - \sum_{k} \Bigg\{ \ln(1-\xi_k) + \frac{\xi_k\,\ln(\xi_k)}{1-\xi_k} \Bigg\} \,. \label{entropy}\ee  For super-Hubble modes and   to leading order in $Y^2$ it follows from the relation (\ref{betakrhor}) that
\be \xi_k = \frac{Y^2}{4} \,\mathcal{F}_+(k,\eta_f) =  \frac{Y^2}{48\pi^2}\Big\{ \ln^2\big(-k\eta_f \big)-2 \ln\big( -k\eta_f\big)\ln\big(-k\eta_0 \big) \Big\} \,.\label{xish} \ee   This is one of the important results in this study.

\section{Power spectrum:}

The power spectrum is given by (\ref{Pofk}) from which it is clear that the phase in (\ref{rhorchifi}) is irrelevant. In terms of the reduced density matrix (\ref{rhorchifi}) we need
\be \int D\chi_f \Big(\chi_{\vk,f}\,\chi_{-\vk,f} \Big)~\rho^r(\chi_f,\widetilde{\chi}_f=\chi_f;\eta_f)\,   = \frac{1}{2\,\alpha_k} =  \frac{1}{2}~ \widetilde{g}_+(k,\eta) \simeq \frac{g_+(k,\eta)}{2}\,\Big[1 + Y^2 \,\mathcal{F}_+(k,\eta_f) +\cdots \Big]  \,,  \label{powchi} \ee where $\mathcal{F}_+$ is given by eqn. (\ref{Fplus}) and we used the normalization (\ref{unitarho}). The secular growth of the correction term $\mathcal{F}_+$ as $-k\eta \rightarrow 0$, leading  eventually to a breakdown of the perturbative expansion can be systematically re-summed via the dynamical renormalization group \cite{drg,nigel}. Following the treatment in ref.\cite{boyfer} we implement this resummation program to obtain a renormalization group improved power spectrum. Returning to the perturbative solution to the equation of motion, eqn.  (\ref{pertsolPsi}), we consider that the zeroth- order solution is given by the growing mode, namely, we take
\be \Psi^c_{\vk,0}(\eta_1) = Q_k\,g_+(k,\eta_1)\,, \label{zerodrg} \ee in eqn.  (\ref{pertsolPsi}), and consider only the term $I[k;\eta_1]$ in the inhomogeneity of the first order equation of motion (\ref{Psi1eom}), because   this term yields the dominant secular growth at long time in the super-Hubble limit. After renormalization and following the steps leading to eqn. (\ref{tilgsfin}) we find
\be \Psi^c_{\vk,1}(\eta ) = Q_k\,g_+(k,\eta )\,\Big[ 1 + Y^2 \,\mathcal{F}_+(k,\eta )+\cdots\Big] \,. \label{Psicgrow} \ee We introduce a (wave-function) renormalization of the amplitude, $\mathcal{Z}[\overline{\eta}]$, and an arbitrary renormalization scale $\overline{\eta}$ to write the amplitude $Q_{\vk}$ as
 \be Q_{\vk} = Q_{\vk}[\overline{\eta}]\mathcal{Z}[\overline{\eta}] ~~;~~\mathcal{Z}[\overline{\eta}] = 1+Y^2\, z_1[\overline{\eta}] + \cdots \,.\label{wfren} \ee Inserting this expansion in the solution (\ref{Psicgrow}), yields
 \be \Psi^c_{\vk}(\eta ) = Q_{\vk}[\overline{\eta}]\,g_+(k,\eta )\,\Big[ 1 + Y^2 \Big( \,\mathcal{F}_+(k,\eta )+z_1[\overline{\eta}]\big) +\cdots\Big] \,. \label{Psicgrow2} \ee The perturbative expansion is improved by choosing the coefficient $z_1[\tau]$ to cancel the secularly growing correction from $\mathcal{F}_+$ at the (arbitrary) scale $\overline{\eta}$, namely
\be \Psi^c_{\vk}(\eta ) = Q_{\vk}[\overline{\eta}]\,g_+(k,\eta )\,\Big[ 1 + Y^2 \Big( \,\mathcal{F}_+(k,\eta )-\mathcal{F}_+(k,\overline{\eta})\big) +\cdots\Big] \,. \label{Psicgrow3} \ee Since the solution $\Psi^c_{\vk},(\eta)$ does not depend on the scale $\overline{\eta}$, it obeys the \emph{dynamical renormalization group equation}\cite{drg,nigel}
\be \frac{d\,\Psi^c_{\vk}(\eta ) }{d\overline{\eta}} = 0\,,  \label{drgeqn}\ee namely, to leading order in $Y^2$
\be \frac{d\,Q_{\vk}[\overline{\eta}] }{d\overline{\eta}} - Y^2\,Q_{\vk}[\overline{\eta}]~\frac{d\, \mathcal{F}_+(k,\overline{\eta}) }{d\overline{\eta}}   = 0\,.  \label{drgeq2}\ee The solution of this equation is
\be Q_{\vk}[\overline{\eta}] = Q_{\vk}[\overline{\eta}^{\,*}]\, \exp\Big[Y^2 \Big(\mathcal{F}_+(k,\overline{\eta})-\mathcal{F}_+(k,\overline{\eta}^{\,*}) \Big)  \Big] \label{drgsolu}\ee We choose the scale $\overline{\eta}^{\,*}$ to correspond to the time at which the mode of wavevector $k$ crosses the Hubble radius, namely $ -k\overline{\eta}^{\,*} = 1$ for two reasons: i) at this time scale the corrections to the mode functions are within the perturbative regime and the amplitude has not changed substantially, ii) at this scale it follows from (\ref{Fplus}) that $\mathcal{F}_+(k,\overline{\eta}^{\,*})=0$. With this physically motivated choice, and now finally setting $\overline{\eta} \equiv \eta$, the renormalization group improved growing solution is
\be \Psi^c_{\vk}(\eta ) = Q_{\vk}\big|_{hc} \, \,g_+(k,\eta )\,\,e^{Y^2 \mathcal{F}_+(k, {\eta})}\,   \label{drgimpsol} \ee where $Q_{\vk}\big|_{hc}\simeq Q_{\vk}$ up to perturbatively small (and non-secular)  corrections in $Y^2$ is the amplitude at ``Hubble-crossing''. The dynamical renormalization group improved solution is equivalent to the solution obtained via the quantum master equation as shown in ref.\cite{boydensmat}.

Replacing this renormalization group improved solution into the analysis of the previous section is tantamount to replacing
\be \widetilde{g}_+(k;,\eta ) \rightarrow g_+(k,\eta )\,\,e^{Y^2 \mathcal{F}_+(k, {\eta})} \,, \label{gtilimp}\ee in all expressions leading to the reduced density matrix. Using the leading order result (\ref{xish}) we obtain the power spectrum for super-Hubble wavelengths at the end of the inflationary era
\be \mathcal{P}(k) = \frac{H^2}{4\pi^2}\,\, e^{8\,\xi_k(\eta_f)}  \,, \label{pofkfini}\ee this result  establishes a direct relationship between the corrections to the power spectrum and the entanglement entropy (\ref{entropy}).

\section{Discussion:}\label{sec:discussion}

Several aspects of the results obtained in the previous sections merit discussion.

\textbf{i:)    } The perturbative argument indicates that the growth of the entanglement entropy is associated with the production of fermion-antifermion pairs which becomes enhanced when  the physical wavelength of the scalar fluctuation becomes super-Hubble. As pair production is enhanced and these degrees of freedom are traced out of the total density matrix to yield the reduced density matrix for the scalar fluctuations, more and more information is lost in coarse graining these degrees of freedom. This information loss is manifest as a growth of entropy\cite{boyinfo}. The effective action confirms this interpretation since the term responsible for the mixing is a consequence of the interactions and fermion pair production.
The growth of entropy for super-Hubble fluctuations has also been found numerically in ref.\cite{hollowood} in a different model with the inflaton coupling to a massless scalar field conformally coupled to gravity\cite{boydensmat,boyeffcosmo1}. Our study thus confirms the growth of entropy upon tracing over ``unobserved'' degrees of freedom providing an analytic description of the entanglement entropy for super-Hubble modes within a very different context of the inflaton Yukawa coupled to fermions.

\vspace{1mm}

\textbf{ii:) } The fact that the entanglement entropy and the corrections to the power spectrum are correlated is understood   from the fact that both are determined by the inflaton self-energy, namely the fermion-antifermion loop. In the perturbative approach this is manifest in the matrix elements of the reduced density matrix, see fig.(\ref{fig:secondorder}) and  in the effective action by the fermionic correlators (\ref{ggreat},\ref{lesser}) which determine the self energy and are obviously given by the loop in fig. (\ref{fig:secondorder}).

At the one loop level the effective action is gaussian, therefore a relationship between the entanglement entropy and the corrections to the inflaton correlator up to order $Y^2$ is expected. What is perhaps unexpected is that the relationship is given by equation (\ref{pofkfini}), namely in the form of a running of the power spectrum.

\vspace{1mm}

\textbf{iii:)} As discussed in detail in ref.\cite{boyfer}, if we restore the \emph{renormalized} mass, keeping $M^2_R(\eta_0) \neq 0$, the power spectrum becomes
\be \mathcal{P}(k) = \frac{H^2}{4\pi^2}\,\, e^{\frac{M^2_R(\eta_0)}{3H^2}\,\ln[-k\eta]}~e^{Y^2\,\mathcal{F}_+(k,\eta)}  \,, \label{pofkMr}\ee the exponent can be combined in the form
\be \Bigg\{ \frac{M^2_R(\eta_0)}{3H^2} - \frac{Y^2}{6\pi^2}\,\ln\big[-k\eta_0 \big] \Bigg\}\,\ln\big[-k\eta\big]+ \frac{Y^2}{12\pi^2}\,\ln^2\big[-k\eta  \big]\,. \label{fullexpo}\ee As is shown in section (\ref{sec:reduced}) (see eqn. (\ref{Mren}) and discussion below it) and in ref.\cite{boyfer},   the term in the bracket is invariant under a change of scale $\eta_0$ thus the total power spectrum is indeed independent of this renormalization scale. However, choosing $\eta_0$ to coincide with the beginning of slow roll and setting the renormalized mass to vanish at this scale leaves the scale $\eta_0$ as a remnant in the power spectrum. This is similar to the emergence of a renormalization scale in renormalized correlation functions that break scale invariance.

\vspace{1mm}

\textbf{iv:)} The correction to the power spectrum and the entanglement entropy are determined by the factor
\be Y^2\mathcal{F}_+(k,\eta_f) = \frac{Y^2}{12\pi^2}\Big\{ \ln^2\big(-k\eta_f \big)-2 \ln\big( -k\eta_f\big)\ln\big(-k\eta_0 \big) \Big\} \,,\ee considering that the total number of efolds $N_T = \ln(\eta_0/\eta_f) \simeq 60$ and that the wavevectors $k$ of cosmological relevance cross the Hubble radius about $10 $ e-folds before the end of inflation at $\eta_f$ it follows that
\be Y^2 \mathcal{F}_+(k,\eta_f) \lesssim  10\,\times \,Y^2  \,, \label{valpt}\ee even considering that Hubble crossing occurs at the beginning of inflation, $-k\,\eta_0 \simeq 1$ (still super-Hubble today) yields
 \be Y^2 \mathcal{F}_+(k,\eta_f) \lesssim  30\,\times \,Y^2  \,. \label{valpt2} \ee
 Therefore with $Y < 10^{-1} $ this contribution is positive and small and perturbation theory is valid. In particular the condition   $0 \leq \lambda_n < 1 $ for the eigenvalues of the normalized (mixed state) reduced density matrix and the expression for the entanglement entropy is fulfilled. Therefore $\lambda_n$ can be safely assigned a probability interpretation for the whole range of wavevectors that cross the Hubble radius between the beginning and end of inflation for moderate Yukawa couplings.

There is a caveat in this argument. In principle allowing the wavevector $k$ to be arbitrarily small it could lead to $Y^2 \mathcal{F}_+(k,\eta_f) > 1$. This possibility results in a breakdown of perturbation theory. While the dynamical renormalization group provides a systematic resummation for the power spectrum, there is no natural manner to extend this well understood resummation framework to the entanglement entropy. Such resummation program for the entanglement entropy remains to be studied further.

\vspace{1mm}

\textbf{v:)} It is convenient to introduce the ``pivot'' scale $k_f = -1/\eta_f$   corresponding to the scale that crosses the Hubble radius at the end of inflation, in terms of which
\be Y^2\mathcal{F}_+(k,\eta_f) =  -\frac{Y^2}{12\pi^2}\Big\{2\,N_T\,\ln(k/k_f) + \ln^2(k/k_f)\Big\} \label{newFplus}\ee with $k\ll k_f$.  The power spectrum (\ref{pofkfini}) can now be written in terms of a correction to the index $\delta n_s$ and  running $\alpha_s$ as
\be \mathcal{P}(k) = \Big(\frac{H}{2\pi} \Big)^2\, \Big( \frac{k}{k_f}\Big)^{\delta n_s + \alpha_s \ln(k/k_f) } \,,\label{newPofk}\ee with
\be \delta n_s = - \frac{N_T\,Y^2}{3\pi^2} ~~;~~ \alpha_s = -\frac{Y^2}{6\pi^2}\,, \label{tiltrun}\ee suggesting a hierarchy $\alpha_s \simeq n_s/N_T;\cdots$.

We note that a change of the ``pivot'' scale $k_f$ results in a change of the overall amplitude \emph{and} a change of $\delta n_s$, these changes have been discussed also in ref.\cite{liddle}.

Therefore we find a correction to the index $n_s$  and a negative running $\alpha_s$  but \emph{not} a running of the running, $\beta_s \simeq 0$ to this order.

\vspace{1mm}

\textbf{vi:)  } A corollary of this study is that even in absence of (scalar) fields that could contribute to entropy perturbations, the coupling of the inflaton to other degrees of freedom that do not contribute \emph{directly} to cosmological perturbations and are  ``traced over'',  lead to entropy production. This entanglement entropy is different from a thermal entropy, but nevertheless imply a loss of information and must be included in the entropy budget both during and post inflation. Thus even without explicit entropy perturbations, the entanglement entropy resulting from particle production contributes to the entropy budget during the inflationary stage.

\vspace{1mm}

\textbf{vii:) Caveats: } We have established a relation between the entanglement entropy and corrections to the power spectrum for \emph{inflaton} fluctuations, \emph{not} for curvature perturbations. The latter are the perturbations relevant for temperature anisotropies, therefore a comparison between the results obtained here and the observational data are not very meaningful. Although our analysis so far does not apply directly to curvature perturbations, it suggests that  the underlying fundamental physical processes, namely self-energy loop corrections of ``unobserved'' spectator fields will lead to similar results for them. This expectation is borne out of the analysis in ref.\cite{kahya1}, that showed the emergence of secular logarithms from loops of ``spectator'' fields.

\section{Conclusions and further questions:}

The main premise of our study is  that the coupling of the inflaton to the degrees of freedom that populate the post-inflationary reheating phase, influence the  dynamics  of the inflaton \emph{during} inflation. We consider the inflaton Yukawa coupled to light fermions, assuming that the scale of inflation is much higher than the electroweak scale. The full density matrix is evolved in time from an initial factorized vacuum state and the fermionic degrees of freedom are traced out of the full density matrix yielding a reduced \emph{mixed} density matrix whose time evolution is determined by a non-equilibrium effective action. A perturbative study of the reduced density matrix reveals that profuse fermion pair production when the wavelengths of the inflaton fluctuations become super-Hubble, result  in   growth of the \emph{entanglement entropy}.

We obtain the one-loop effective action which confirms  that the fermionic self-energy leads to secular growth of inflaton correlations \emph{and} the entanglement entropy. The entanglement entropy is a manifestation of the \emph{information loss} in the effective field theory\cite{boyinfo}, arising from tracing over the ``unobserved'' degrees of freedom. As more fermion pairs are produced, tracing these degrees of freedom out of the density matrix implies more information loss and a concomitant growth in the entanglement entropy.

 We establish a direct relation between scale invariance violations of the inflaton power spectrum and the entanglement entropy, $\mathcal{P}(k) = \mathcal{P}_0(k)\,\,\exp\{8\,\xi_k\}$ with $\mathcal{P}_0(k)$ the unperturbed (scale invariant) power spectrum and Von-Neumann entanglement entropy   $S_{vN} = - \sum_{k} \Big[ \ln(1-\xi_k) + \frac{\xi_k\,\ln(\xi_k)}{1-\xi_k} \Big]$. For super-Hubble fluctuations we find
$\xi_k = -\frac{Y^2}{48\pi^2}\Big\{2\,N_T\,\ln(k/k_f) + \ln^2(k/k_f)\Big\}$ with $Y$ the Yukawa coupling, $N_T$ the total number of e-folds during inflation, and $k_f$ a ``pivot'' scale corresponding to the mode that crosses the Hubble radius at the end of inflation.  The correction to the index and its running are given by

\be \delta n_s = - \frac{N_T\,Y^2}{3\pi^2} ~~;~~ \alpha_s = -\frac{Y^2}{6\pi^2}\,, \ee with vanshing running of the running $\beta_s \simeq 0$ to the order considered.

A corollary of our study is that even in absence of scalar entropy (isocurvature) perturbations, the coupling to the inflaton to degrees of freedom that are not directly observed and are integrated out into an effective action contribute to entropy production during the inflationary stage as a consequence of the production of correlated pairs. This entropy, different from the thermal variety, is \emph{imprinted} on the power spectrum of fluctuations and  must be included in  the cosmological entropy budget.

\vspace{1mm}

\textbf{Further questions:}

In this article we focused on studying the influence of ``unobserved'' degrees of freedom upon the inflaton. It remains to understand how to implement the formulation presented here to curvature perturbations, in particular addressing the important issue of gauge invariance. Therefore, while the results obtained here are indicative of the effect of degrees of freedom that are traced over, and   the physical reasons (self-energy corrections from particle production) for the correlation between entanglement entropy and corrections to (near) scale invariance are clear, such relation for curvature perturbations must be studied further.

Although the relation between the entanglement entropy and scaling violations of the power spectrum is fundamentally important as a characterization of the corrections from ``unobserved'' degrees of freedom, it is not clear whether these corrections will be observationally distinguishable from those of ``ordinary'' slow roll, or other sources such as isocurvature perturbations. Thus the observational consequences of the results in this study   highlighting the influence of degrees of freedom that do not directly seed curvature perturbations and the relation with entanglement entropy production remain to be studied further.

It would be very tantalizing if CMB observations can discriminate between scaling violations in the form of corrections to the tilt and running induced by ``unobserved'' degrees of freedom from those predicted by slow roll. Such observation may open the window to glean other degrees of freedom beyond the inflaton in inflationary cosmology.

\acknowledgements The author   gratefully   acknowledges  support from NSF through grant PHY-1506912.

\appendix
\section{Coefficients in  eqn. (\ref{Seffulti2})}\label{app:coeffs}

In terms of the real mode functions  $g_{\pm}(k,\eta_0)$ the coefficients $\Omega_{R,k}, \Omega_{I,k} $ in eqn. (\ref{omes})  are given by
\bea \frac{1}{\Omega_{R,k}} & = & g^2_+(k,\eta_0)+ g^2_-(k,\eta_0) \label{OmegaRzero}\\
 \frac{\Omega_{I,k}}{\Omega_{R,k}} & = & -\Big[ g^{'}_+(k,\eta_0)\,g_+(k,\eta_0)+ g^{'}_-(k,\eta_0)\,g_-(k,\eta_0)\Big] \,. \label{ratiozero}\eea
 The remaining coefficients in eqn. (\ref{Seffulti2}) are given by the following expressions,

\be A_{k,i} = \Bigg\{ \frac{d}{d\eta_1}\,\Bigg[\frac{\widetilde{D}_k[\eta_f;\eta_1]}{\widetilde{D}_k[\eta_f;\eta_0]}\Bigg] \Bigg\}_{\eta_1=\eta_0} ~~;~~ A_{k,f} = \Bigg\{ \frac{d}{d\eta_1}\,\Bigg[\frac{\widetilde{D}_k[\eta_f;\eta_1]}{\widetilde{D}_k[\eta_f;\eta_0]} \Bigg]  \Bigg\}_{\eta_1=\eta_f} \,,\label{Aif} \ee
\be B_{k,i} = \Bigg\{ \frac{d}{d\eta_1}\,\Bigg[\frac{\widetilde{D}_k[\eta_1;\eta_0]}{\widetilde{D}_k[\eta_f;\eta_0]}\Bigg] \Bigg\}_{\eta_1=\eta_0} ~~;~~ B_{k,f} = \Bigg\{ \frac{d}{d\eta_1}\,\Bigg[\frac{\widetilde{D}_k[\eta_1;\eta_0]}{\widetilde{D}_k[\eta_f;\eta_0]} \Bigg]  \Bigg\}_{\eta_1=\eta_f} \,,\label{Bif} \ee
\be C_{k,i} = \Bigg\{ \frac{d}{d\eta_1}\,H_1(k,\eta_1) \Bigg\}_{\eta_1=\eta_0}~~;~~ C_{k,f} = \Bigg\{ \frac{d}{d\eta_1}\,H_1(k,\eta_1) \Bigg\}_{\eta_1=\eta_f}\,,  \label{Cif}\ee
\be D_{k,i} = \Bigg\{ \frac{d}{d\eta_1}\,H_2(k,\eta_1) \Bigg\}_{\eta_1=\eta_0}~~;~~ D_{k,f} = \Bigg\{ \frac{d}{d\eta_1}\,H_2(k,\eta_1) \Bigg\}_{\eta_1=\eta_f}\,, \label{Dif}\ee where
$H_{1,2}(k,\eta_1)$ are given by (\ref{Hfinas}) with
\be h_1(k,\eta_1) = - \int^{\eta_1}_{\eta_0} d\eta_2 \int^{\eta_f}_{\eta_0} d\eta_3 ~ D_k[\eta_1,\eta_2]\,N_k(\eta_2,\eta_3)\,\frac{D_k[\eta_f,\eta_3]}{D_k[\eta_f,\eta_0]} \,,\label{h1}\ee
\be h_2(k,\eta_1) = - \int^{\eta_1}_{\eta_0} d\eta_2 \int^{\eta_f}_{\eta_0} d\eta_3 ~ D_k[\eta_1,\eta_2]\,N_k(\eta_2,\eta_3)\,\frac{D_k[\eta_3,\eta_0]}{D_k[\eta_f,\eta_0]} \,,\label{h2}\ee
\be J_{1,k} =   \int^{\eta_f}_{\eta_0} d\eta_1 \int^{\eta_f}_{\eta_0} d\eta_2 ~ \frac{D_k[\eta_f,\eta_1]}{D_k[\eta_f,\eta_0]}\,N_k(\eta_1,\eta_2)\,\frac{D_k[\eta_f,\eta_2]}{D_k[\eta_f,\eta_0]} \,,\label{j1}\ee
\be J_{2,k} =   \int^{\eta_f}_{\eta_0} d\eta_1 \int^{\eta_f}_{\eta_0} d\eta_2 ~ \frac{D_k[\eta_1,\eta_0]}{D_k[\eta_f,\eta_0]}\,N_k(\eta_1,\eta_2)\,\frac{D_k[\eta_2,\eta_0]}{D_k[\eta_f,\eta_0]} \,,\label{j2}\ee
\be J_{3,k} =   \int^{\eta_f}_{\eta_0} d\eta_1 \int^{\eta_f}_{\eta_0} d\eta_2 ~ \frac{D_k[\eta_f,\eta_1]}{D_k[\eta_f,\eta_0]}\,N_k(\eta_1,\eta_2)\,\frac{D_k[\eta_2,\eta_0]}{D_k[\eta_f,\eta_0]} \,.\label{j3}\ee

\section{Coefficients for $Y=0$.}\label{app:yzero}

In this case only the coefficients $A,B$ in eqn. (\ref{Seffulti2}) are different from zero. From (\ref{Aif}) and (\ref{Bif}) with
\be \widetilde{D}_k[\eta_1;\eta_2]=  {D}_k[\eta_1;\eta_2]= {g}_+(k,\eta_1)\, {g}_-(k;\eta_2)- {g}_+(k;\eta_2)\, {g}_-(k,\eta_1)\,, \ee we find
\bea A_{k,i} & = & \frac{d_i}{{D}_k[\eta_f;\eta_0]} \label{Aiyzero}\\
A_{k,f} & = & \frac{1}{{D}_k[\eta_f;\eta_0]} \label{Afyzero} \\
B_{k,i} & = & -\frac{1}{{D}_k[\eta_f;\eta_0]} \label{Biyzero} \\
B_{k,f} & = &  \frac{d_f}{{D}_k[\eta_f;\eta_0]} \,,\label{Bfyzero} \eea where
\bea d_i & = &  {g}_+(k,\eta_f)\, {g}'_-(k;\eta_0)- {g}'_+(k;\eta_0)\, {g}_-(k,\eta_f) \label{diyzero}\\ d_f & = &  {g}'_+(k,\eta_f)\, {g}_-(k;\eta_0)- {g}_+(k;\eta_0)\, {g}'_-(k,\eta_f)\,. \label{dfyzero} \eea

\section{Analysis of   coefficients for $Y\neq 0$ in the super-Hubble limit.}\label{app:shcoeffs}
For this analysis we consider that the renormalized mass of the inflaton field is $M_R \ll H$ therefore taking $\nu_\chi =3/2$ in the mode functions. The coefficient $B_{k,f}$ only contributes to the \emph{phase} of the reduced density matrix (\ref{Seffulti2}), namely the coefficient $\gamma_k$ in eqns. (\ref{rhorfi},\ref{rhorchifi}), therefore is not relevant for either the power spectrum or the entanglement entropy as discussed in the text.  The main ingredients in the analysis of the super-Hubble limit $-k\eta \rightarrow 0^+$ are:
\be   g_+(k,\eta) =  \frac{1}{k^{3/2}\,\eta} ~~;~~ g_-(k,\eta) =  \frac{1}{3}\,k^{3/2}\,\eta^2 \,, \ee
\be \widetilde{g}_\pm(k,\eta_f)   =    g_\pm(k,\eta_f)\Big[ 1 + Y^2 \,\mathcal{F}_\pm(k,\eta_f)\Big] \ee with
\bea \mathcal{F}_+(k,\eta_f) & = & \frac{1}{12\pi^2}\Big\{ \ln^2\big(-k\eta_f \big)-2 \ln\big( -k\eta_f\big)\ln\big(-k\eta_0 \big) \Big\}+\cdots  \\
\mathcal{F}_-(k,\eta_f) & = & \mathrm{finite}\,\,\mathrm{constant}\,.  \eea Therefore, it follows that
\be D_k[\eta_f,\eta_0] \simeq  g_+(k,\eta_f) \,g_-(k,\eta_0) ~~;~~ \widetilde{D}_k[\eta_f,\eta_0] \simeq \widetilde{g}_+(k,\eta_f)\,g_-(k,\eta_0)\,. \ee Furthermore, from eqn. (\ref{tilgszero}) and the Wronskian condition (\ref{Wronk}) we find
\be B_{k,i} = -\big(\widetilde{D}_k[\eta_f,\eta_0]\big)^{-1} ~~;~~
  A_{k,f} \simeq \Big[1+Y^2\,\mathcal{F}_+(k,\eta)\Big]\,\big(\widetilde{D}_k[\eta_f,\eta_0]\big)^{-1} \,,\label{ABkfinal} \ee and

  \be A_{k,i}= \frac{g'_-(k,\eta_0)}{g_-(k,\eta_0)} \simeq \mathcal{O}(1) \,. \ee   The super-Hubble limit of the kernel $N_k(\eta_1,\eta_2)$ given by eqn. (\ref{noisiker}) is dominated by the delta function. Therefore for the terms that involve the nested integrals of $N_k(\eta_1,\eta_2)$ we integrate by parts,    the contribution from the upper limits vanish because $D[\eta,\eta]=0$, and that of the lower limit yields a perturbatively small correction without secular logarithms. In the integrands we take $D[\eta_f,\eta] \simeq g_+(k,\eta_f) \,g_-(k,\eta)$.

   To leading order we find
  \bea && J_{a,k} \simeq Y^2\,\times \,\mathrm{finite}~\mathrm{constant} ~~;~~ a=1,2,3 \nonumber \\
  &&  C_{i,k} ~,~ D_{i,k} \simeq Y^2\,\times \,\mathrm{finite}~\mathrm{constant} \eea

  Up to $\mathcal{O}(Y^2)$ in the squared bracket of (\ref{betak}) we need the products
  \be J_{3,k}\,A^{(0)}_{k,f}~,~D_{k,i}\,A^{(0)}_{k,f}~,~C_{k,f}\,A^{(0)}_{k,f} ~~;~~ A^{(0)}_{k,f} =\big(\widetilde{D}_k[\eta_f,\eta_0]\big)^{-1}\,, \ee from the above results we find that these products yield terms of the form
  \be Y^2 \, \big(\widetilde{D}_k[\eta_f,\eta_0]\big)^{-1} \ee which are subleading compared to the $Y^2\,\ln^2(-k\eta)\,\big(\widetilde{D}_k[\eta_f,\eta_0]\big)^{-1}$ terms in $A_{k,f}$.

  Terms of the form
  \be Y^2 ~ (\mathrm{constant}) ~\big(\widetilde{D}_k[\eta_f,\eta_0]\big)^{-1} ~,~  Y^2 \big(\widetilde{D}_k[\eta_f,\eta_0]\big)^{-2} \propto Y^2 ~ ( k \eta_f)  \, \big(\widetilde{D}_k[\eta_f,\eta_0]\big)^{-1} \,, \ee   from the coefficients $C,D$ in (\ref{omek}) are subleading perturbative corrections as compared to the terms of $\mathcal{O}(1)$  and can be safely neglected.

  Therefore up to leading order in $Y^2$ and secular logarithmic terms we find
  \be \omega_k = \frac{1}{4\,\Omega_{R,k}} \,\Big[\Omega^2_{R,k}+ (A_{k,i}+\Omega_{I,k})^2 \Big] \,, \label{omegakfinal}\ee
  \be \alpha_k  = \frac{1}{4\omega_k\, \big(\widetilde{D}_k[\eta_f,\eta_0]\big)^{2}} = \frac{1}{\widetilde{g}^{\,2}_+(k,\eta_f)}  \,,\label{alfakfinal}\ee and
  \be \beta_k = \frac{A^2_{k,f}}{4\,\Omega_{R,k}}\Big[ \frac{(A_{k,i}+\Omega_{I,k})^2}{4\,\omega_k\,\Omega_{R,k} }-1 \Big] = \frac{1}{4} \,\frac{A^2_{k,f}}{4\,\omega_{k}} = \frac{\alpha_k}{4}\,\Big[1+Y^2\,\mathcal{F}_+(k,\eta)\Big] \,.\label{betakfinal}\ee


\begin{thebibliography}{99}

\bibitem{wmap} C. L. Bennett \emph{et.al. WMAP collaboration} Astrophys.J.Suppl. \textbf{208},    20, (2013).

  \bibitem{spergel} D. N. Spergel \emph{et.al. WMAP collaboration} Astrophys.J.Suppl. \textbf{148}, 175 (2003).

\bibitem{planck} P.A.R. de Ade \emph{et. al. Planck collaboration} Astron.Astrophys. \textbf{594}, A20 (2016); \textbf{594}, A13 (2016).

 \bibitem{guth} A. H. Guth, Phys. Rev. \textbf{D23}, 347 (1981).

 \bibitem{linde} A. Linde, Phys. Lett. \textbf{B108}, 389 (1982).



    \bibitem{mukhanov} V. F. Mukhanov, G. V. Chivisov, JETP Lett. \textbf{33}, 532 (1981).

 \bibitem{bran} V. Mukhanov, H. Feldman and R. Brandenberger, Phys. Rep. \textbf{215}, 203 (1992).





 \bibitem{reheat1} A. Dolgov and D. Kirilova, Sov. J. Nucl. Phys. \textbf{51} (1990) 172;
  J. H. Traschen and R. H. Brandenberger, Phys. Rev. \textbf{D42} (1990) 2491;
  L. Kofman, A. D. Linde and A. A. Starobinsky, Phys. Rev. Lett. \textbf{73}(1994) 3195; L. Kofman, A. D. Linde and A. A. Starobinsky, Phys. Rev. \textbf{D56} (1997) 3258


   \bibitem{reheatrev} For a recent comprehensive review on reheating see:  M. A. Amin, M. P. Hertzberg, D. I. Kaiser, J. Karouby,  Int. J. Mod. Phys. \textbf{D 24}, 1530003 (2015), and references therein.

    \bibitem{woodardcosmo} S. P. Miao, N. C. Tsamis, R. P. Woodard, arXiv:1002.4037; R. P. Woodard, arXiv:astro-ph/0310757; T. M. Janssen, S. P. Miao, T. Prokopec, R. P. Woodard,  Class.Quant.Grav.\textbf{25}, 245013 (2008); N. C. Tsamis and R. P. Woodard,   Phys. Lett. \textbf{B
301}, 351 (1993) 351; N. C. Tsamis and R. P. Woodard,   Annals
Phys. \textbf{238},1  (1995); N. C. Tsamis and R. P. Woodard,   Phys.
Rev. \textbf{D 78}, 028501 (2008).

   \bibitem{proko1} G. Lazzari, T. Prokopec,   arXiv:1304.0404;    J. Weenink, T. Prokopec,  	arXiv:1108.3994; J. F. Koksma, T. Prokopec, M. G. Schmidt, Phys.Rev. \textbf{D81} 065030 (2010); D. Glavan, T. Prokopec, V. Prymidis,  Phys. Rev. \textbf{D 89}, 024024 (2014);
         A. Marunovic, T. Prokopec,  Phys.Rev.\textbf{D83}, 104039 (2011).



  \bibitem{decayds}  D. Boyanovsky, H. J. de Vega, Phys.~Rev. \textbf{D70},  063508  (2004);  D. Boyanovsky, H. J. de Vega, N. G. Sanchez,  Phys.~Rev.\textbf{D71} 023509 (2005); Nucl.~Phys. \textbf{B747}, 25 (2006).

 \bibitem{akhmedov} E. T. Akhmedov, A. Roura, A. Sadofyev, Phys.~Rev.\textbf{D82}, 044035 (2010);  E. T. Akhmedov, P. V. Buividovich, Phys.~Rev.\textbf{D78}, 104005 (2008);  E. T. Akhmedov, Mod.Phys.Lett.\textbf{A25},2815 (2010); E. T. Akhmedov, P. V. Buividovich, D. A. Singleton, Phys.Atom.Nucl. \textbf{75 }, 525 (2012); E. T. Akhmedov, JHEP \textbf{1201}, 066 (2012);  E. T. Akhmedov Int. Jour. of Mod. Phys. \textbf{D23}, 1430001  (2014).


\bibitem{woodard1} N. C. Tsamis, A. Tzetzias, R. P. Woodard,  JCAP \textbf{1009}, 016 (2010); N. C. Tsamis, R. P. Woodard, Nucl.Phys. \textbf{B724}, 295 (2005);   R. P. Woodard,  arXiv:astro-ph/0502556.

\bibitem{prokowood} T. Prokopec, N. C. Tsamis, R. P. Woodard, Ann. of Phys. \textbf{323}, 1324 (2008).

\bibitem{onemli} V.K. Onemli, Phys. Rev. \textbf{D 89}, 083537 (2014) ;  Phys. Rev.\textbf{ D 91}, 103537 (2015); arXiv: 1510.02272.

\bibitem{sloth1} S. B. Giddings, M. S. Sloth, JCAP \textbf{1101}, 023 (2011).

\bibitem{sloth2} S. B. Giddings, M. S. Sloth,   JCAP \textbf{1007}, 015 (2010); S. B. Giddings, M. S. Sloth, Phys.Rev. \textbf{D84} (2011) 063528; R. Kumar Jain, M. Sandora, M. S. Sloth,  JCAP \textbf{1506}, 016 (2015).

\bibitem{riotto} A. Riotto and M. S. Sloth, JCAP \textbf{0804}, 030 (2008); A. Kehagias, A. Riotto, 	Nucl. Phys.  \textbf{B868},  577 (2013); Nucl. Phys. \textbf{B864},   492 (2012); A. Riotto, M. S. Sloth, JCAP \textbf{10},  003 (2011).

\bibitem{fermionswoodpro}  T. Prokopec and R. P. Woodard, JHEP \textbf{0310}, 059
(2003); B. Garbrecht and T. Prokopec, Phys. Rev. \textbf{D 73}, 064036 (2006); S. P. Miao and R. P. Woodard,  	Class. Quant. Grav.\textbf{23},1721 (2006); S. P. Miao and R. P. Woodard,   Phys. Rev. \textbf{D 74}, 044019
(2006).

 \bibitem{picon} C. Armendariz-Picon,  	JCAP \textbf{0702} 031 (2007).

 \bibitem{lello} L. Lello, D. Boyanovsky, R. Holman,   JHEP \textbf{04},055 (2014).

 \bibitem{rich} H. Collins, R. Holman, A. Ross, JHEP 1302 (2013) 108; C.P. Burgess, R. Holman, L. Leblond, S. Shandera,
 JCAP \textbf{1010}, 017 (2010); C.P. Burgess, L. Leblond, R. Holman, S. Shandera,  JCAP \textbf{1003}, 033 (2010).

 \bibitem{raja} J. Kumar, L.Leblond, A. Rajaraman, JCAP \textbf{1004}, 024 (2010); D. P. Jatkar, L. Leblond, A. Rajaraman, Phys.Rev. \textbf{D85}, 024047  (2012); A. Rajaraman, Phys.Rev.\textbf{D82}, 123522 (2010);   Int.J.Mod.Phys. \textbf{A30},1550173 (2015).

 \bibitem{richboy} D. Boyanovsky, R. Holman, JHEP, \textbf{2011}, 47 (2011).

 \bibitem{boyan} D. Boyanovsky,  Phys. Rev. \textbf{D 86}, 023509 (2012);  Phys. Rev. \textbf{D 85}, 123525 (2012).

\bibitem{serreau} J. Serreau, Phys.Lett. \textbf{B728}, 380 (2014); F. Gautier, J. Serreau; Phys.Lett. \textbf{B727}, 541 (2013); J. Serreau, R. Parentani, Phys.Rev. \textbf{D87}, 085012 (2013); M. Guilleux, J. Serreau,  	Phys. Rev. \textbf{D 92}, 084010 (2015).

\bibitem{parentani} R. Parentani, J. Serreau, Phys.Rev. \textbf{D87}  045020, (2013).

\bibitem{smit} M. van der Meulen and J. Smit, JCAP \textbf{0711}, 023 (2007).


    \bibitem{polyakov} A. M. Polyakov, Nucl.Phys.\textbf{B834}, 316 (2010); D. Krotov, A. M.

\bibitem{kahya1} E. O. Kahya, V. K. Onemli, R. P. Woodard, Phys.Lett. \textbf{ B694},101 (2010).

\bibitem{kahya2}  S. Boran, E. O. Kahya,  Phys. Rev. \textbf{D 97}, 043507 (2018).

\bibitem{covi} L. Covi, S. Dresti, arXiv:1803.02351

\bibitem{breuer} N. P. Breuer, F. Petruccione, \emph{The theory of open quantum systems}, Oxford University Press, Oxford, 2007.

\bibitem{zoeller} C. Gardiner, P. Zoeller, \emph{Quantum Noise} Springer-Verlag, Berlin (2010).

\bibitem{boyeff} D. Boyanovsky,   New J. Phys. 17 (2015) 063017.

\bibitem{burhol} C.P. Burgess, J. Cline, R. Holman, JCAP \textbf{0310}, 004 (2003),  C. P. Burgess, R. Holman and D. Hoover,  Phys. Rev. \textbf{D 77},063534 (2008); C.P. Burgess, R. Holman, G. Tasinato, M. Williams,   JHEP \textbf{1503}, 090 (2015); C.P. Burgess, R. Holman, G. Tasinato, JHEP \textbf{1601}, 153 (2016)

   \bibitem{boydensmat} D. Boyanovsky,   Phys. Rev. \textbf{D 92}, 023527 (2015).

   \bibitem{hollowood} T. J. Hollowood, J. I. McDonald, Phys. Rev.
    \textbf{D 95}, 103521.

    \bibitem{oshita} N. Oshita,  	Phys. Rev. \textbf{D 97}, 023510 (2018).

    \bibitem{shandera} S. Shandera, N. Agarwal, A. Kamal, arXiv:1708.00493.

    \bibitem{vennin} J. Martin, V. Vennin,   JCAP \textbf{1805}, 05 (2018).

    \bibitem{kanno} S. Kanno, JCAP \textbf{1407}, 029  (2014);
 Phys.Lett. \textbf{B751}, 316 (2015).

    \bibitem{boyeffcosmo1} D. Boyanovsky, Phys. Rev. \textbf{D 93}, 043501 (2016).

\bibitem{boyfer} D. Boyanovsky,  Phys. Rev. \textbf{D 93}, 083507 (2016).



\bibitem{escudero} M. Escudero, H. Ramírez, L. Boubekeur, E. Giusarma, O. Mena,  JCAP \textbf{1602} no.02, 020 (2016).

 \bibitem{cabass} G. Cabass, E. Di Valentino, A. Melchiorri, E. Pajer, J. Silk,  	Phys. Rev. \textbf{D 94}, 023523 (2016).

     \bibitem{kamion} J. Munoz, E. D. Kovetz, A. Raccanelli, M. Kamionkowski, J. Silk, JCAP \textbf{1705}, 032 (2017).

 \bibitem{bruck} C. van de Bruck, C. Longden,  	Phys. Rev. \textbf{D 94}, 021301 (2016).


 \bibitem{kimmo} K. Kainulainen, J. Leskinen, S. Nurmi, T. Takahashi        JCAP \textbf{11}, 002 (2017).

  \bibitem{boyinfo} D. Boyanovsky,  Phys. Rev. \textbf{D 97}, 065008 (2018).


\bibitem{weinbergbook} S. Weinberg, \emph{Gravitation and Cosmology:
principles and applications of the general
theory of relativity. } John Wiley and sons, N.Y. 1972.
\bibitem{BD} N. D. Birrell and P. C. W. Davies, \emph{Quantum fields in curved
space}, Cambridge Monographs in Mathematical Physics, Cambridge
University Press, Cambridge, 1982.
\bibitem{duncan} A. Duncan, Phys. Rev. \textbf{D17}, 964 (1978).
\bibitem{casta} M. A. Castagnino, L. Chimento, D. D. Harari and C.
Nunez, J. Math. Phys. \textbf{25}, 360 (1984).





\bibitem{boydVS} D. Boyanovsky, H. J. de Vega, N. G. Sanchez,  Phys.Rev.\textbf{D72}, 103006 (2005).

\bibitem{baacke} J. Baacke, C. Patzold,  	Phys.Rev. \textbf{D62}, 084008 (2000).

\bibitem{polarski} D. Polarski, A. A. Starobinsky, Class.Quant.Grav.\textbf{13}, 377 (1996); C. Kiefer, D. Polarski, A.A. Starobinsky,  Int.J.Mod.Phys.\textbf{ D7},455  (1998); C. Kiefer, D. Polarski,  Annalen Phys. \textbf{7}, 137 (1998); C. Kiefer, J. Lesgourgues, D. Polarski, A. A. Starobinsky,  Class.Quant.Grav.\textbf{15}:L67-L72 (1998); C. Kiefer, D. Polarski,  Adv.Sci.Lett.\textbf{2}:164 (2009).

    \bibitem{schwinger} J. Schwinger, J. Math. Phys. \textbf{2}, 407
(1961).

    \bibitem{keldysh} L. Keldysh,  Sov. Phys. JETP
\textbf{20}, 1018 (1965).

 \bibitem{maha} P. M. Bakshi and K. T. Mahanthappa,  J.Math.Phys. \textbf{4}1 (1963), J.Math.Phys.\textbf{ 4} 12 (1963).





     \bibitem{bombelli} L. Bombelli, R. K. Koul, J. Lee, R. D. Sorkin, Phys. Rev. \textbf{D34}, 373 (1986).

     \bibitem{srednicki} M. Srednicki, Phys. Rev. Lett. \textbf{71}, 666 (1993).

\bibitem{callan} C. Callan, F. Wilczek, Phys. Lett. \textbf{B333}, 55, (1994).

\bibitem{cardy} P. Calabrese, J. Cardy,  J.Stat.Mech. 0406 (2004) P06002.

\bibitem{neu} Th. M. Nieuwenhuizen, A. E. Allahverdyan,  Phys. Rev. \textbf{E 66}, 036102 (2002).

 \bibitem{feynman} R. P. Feynman, \emph{Statistical Mechanics, A set of Lectures}, (Addison Wesley, California, 1972).

    \bibitem{drg} D. Boyanovsky, H. J. de Vega,  Annals Phys. \textbf{307}, 335 (2003); D. Boyanovsky, H. J. de Vega, S.-Y. Wang, Phys.Rev.\textbf{D67}, 065022 (2003); D. Boyanovsky, H. J. de Vega, Phys.Rev. \textbf{D70},063508  (2004).

\bibitem{nigel} L.-Y. Chen, N. Goldenfeld and Y. Oono, Phys. Rev. Lett. 73, 1311 (1994); Phys. Rev. E 54, 376 (1996).

    \bibitem{liddle} M. Cortês, A. R. Liddle, P. Mukherjee,  	Phys.Rev.\textbf{D75}, 083520 (2007).

\end{thebibliography}
\end{document}